\documentclass[draft]{agujournal2019}
\usepackage{url} 
\usepackage{lineno}
\usepackage{amsmath,amsfonts}
\usepackage{tabularx}
\usepackage{subcaption}

\usepackage{multirow}
\usepackage{setspace}
\usepackage[inline]{trackchanges} 
\usepackage{soul}
\draftfalse

\journalname{Journal of Geophysical Research: Solid Earth}
\begin{document}
\justify

\title{MS{\_}ATpV-FWI: Full Waveform Inversion based on Multi-scale Structural Similarity Index Measure and Anisotropic Total p-Variation Regularization}

%
%




\authors{Liangsheng He\affil{1}, Chao Song\affil{1}, and Cai Liu\affil{1}}

\affiliation{1}{College of Geo-Exploration Science and Technology, Jilin University, Changchun, Jilin
130026, China}




\correspondingauthor{Chao Song}{chao.song@kaust.edu.sa}

\begin{keypoints}
\item MS{\_}ATpV-FWI uses multi-scale structural similarity index measure to extract the characteristics, reducing the risk of cycle skipping

\item MS{\_}ATpV-FWI introduces anisotropic total p-variation regularization to suppress artifacts in the gradient and improve the continuity

\item MS{\_}ATpV-FWI demonstrates its stable and accurate inversion performance with synthetic and field seismic data
\end{keypoints}

%
%

%
%


\begin{abstract}

Full waveform inversion (FWI) is a high-resolution seismic inversion technique popularly used in oil and gas exploration. Traditional FWI employs the $l_2$ norm measurement to minimize the misfit between observed and predicted seismic data. However, when the background velocity is inaccurate or the seismic data lacks low-frequency components, the conventional FWI suffers from cycle skipping, leading to inaccurate inversion results. This paper introduces a multiscale structural similarity index measure (M-SSIM) objective function for FWI. We also incorporate anisotropic total p-variation regularization (ATpV) to further improve the accuracy of FWI. M-SSIM extracts multi-scale structural features of seismic data in terms of both phase and amplitude. These features can reduce the risk of cycle skipping and improve the stability of FWI. Additionally, ATpV applies structural constraints to the velocity gradients, which helps suppress artifacts and preserve the sharp boundaries of geological formations. We propose to use the automatic differentiation (AD) to efficiently and stably optimize this novelly introduced FWI objective function. Both synthetic and field seismic data demonstrate that the proposed method accurately characterizes complex subsurface velocity structures, even when the background velocity is crude, the data lacks low-frequency components, or contains noise. 
\end{abstract}

\section{Introduction}

As oil and gas exploration increasingly targets complex and hidden reservoirs, obtaining high-precision, high-resolution imaging and inversion of intricate subsurface geological structures has become a significant challenge \cite[]{virieuxOverviewFullwaveformInversion2009}. Full waveform inversion (FWI), based on the wave equation, utilizes the full waveform information in seismic data to accurately estimate complex subsurface structures \cite[]{tarantolaInversionSeismicReflection1984}. It can provide high-resolution and high-accuracy subsurface structure imaging results for oil and gas exploration. Traditional FWI uses the $l_2$ norm as the misfit function to minimize the discrepancy between predicted and observed data \cite[]{tarantolaInversionSeismicReflection1984, wang2009reflection, warner2013anisotropic, chi2015correlation, wang2019full}. However, due to the high nonlinearity of FWI, traditional FWI is prone to cycle skipping and may converge to a local minimum when seismic data lacks low-frequency information or the initial model is inaccurate \cite[]{wangReweightedVariationalFullwaveform2023, yangFWIGANFullWaveformInversion2023, duPhysicsInformedRobustImplicit2024}.

To address the limitations of the traditional FWI, some researchers have supplemented seismic data with low-frequency information through frequency extrapolation techniques \cite[]{plessix2010application, liFullwaveformInversionExtrapolated2016, wang2019retrieving, ovcharenko2019deep, sun2021deep}. \citeA{chi2014full} and \citeA{wuSeismicEnvelopeInversion2014} demonstrated that the envelope of seismic data produces artificial ultra-low-frequency information and proposed an envelope inversion (EI) method, which recovers low-wavenumber components of the subsurface model. \citeA{liFullwaveformInversionExtrapolated2016} applied phase tracking to estimate low-frequency components from seismic data, presenting an FWI method based on extrapolated low-frequency data. 

Other researchers have focused on directly estimating an accurate background velocity model for FWI \cite[]{luoFulltraveltimeInversion2016, hondori2015new, datta2016estimating, wang2018building}. \citeA{maWaveequationReflectionTraveltime2013} used dynamic image warping to minimize the time offset between predicted and observed seismic data, updating the low-wavenumber components of the velocity model, and then used traditional FWI to refine the high-wavenumber components for high-resolution inversion. \citeA{sava2004wave}, \citeA{symes2008migration}, and \citeA{alaliEffectivenessPseudoinverseExtended2020} applied migration velocity analysis to obtain a good background velocity, which improved the accuracy of FWI. \citeA{xu2012full} and \citeA{yaoReviewReflectionwaveformInversion2020} used reflection waveform inversion (RWI) to update the background velocity, achieving improved accuracy in deep-layer inversion results. \citeA{wangFrequencydomainWaveequationTraveltime2021} performed wave equation travel time inversion in the frequency domain to obtain a reliable background velocity model, with a sequential implementation of traditional FWI to refine the resolution of the subsurface velocity.

These methods can improve the performance of FWI by supplementing low-frequency data or providing an accurate background velocity model. However, these strategies require two-step inversion processes essentially, which introduces additional computation burdens.

By improving the objective functions of FWI, we can directly reduce the risk of cycle skipping \cite[]{yangApplicationOptimalTransport2018, liuRobustFullwaveformInversion2023, yangWassersteinDistanceBasedFullWaveform2023, wangFWIWFRobustFullwaveform2024}. \citeA{vanleeuwenMitigatingLocalMinima2013} introduced a wave equation penalty term to the traditional $l_2$ misfit function, formulating a wavefield reconstruction inversion method that expands the search space for the global optimal solution and reduces cycle skipping. \citeA{warnerAdaptiveWaveformInversion2016} designed an adaptive waveform inversion method by transforming the minimization of the misfit between observed and predicted data into the minimization of the misfit between the Wiener filter and zero-lag delta functions. \citeA{engquistOptimalTransportSeismic2016}, based on optimal transport theory, proposed the W$_2$-FWI method by applying the Wasserstein-2 (W$_2$) metric, which not only measures amplitude differences but also compares global phase, transforming the traditional direct data misfit measurement into the minimization of transmission cost. \citeA{metivierMeasuringMisfitSeismograms2016} refined this approach with the W$_1$-FWI method using the Wasserstein-1 (W$_1$) metric, reducing cycle skipping. However, these methods have high computational complexity and do not consider the multi-scale structural characteristics of seismic data in terms of both time and space \cite[]{krebs2009fast}.

In this study, we introduce a novel FWI method based on multi-scale structural similarity index measure (M-SSIM) and anisotropic total p-variation regularization (ATpV), referred to as MS{\_}ATpV-FWI. The method constructs a misfit function using M-SSIM, which considers the multi-scale feature discrepancies between predicted and observed seismic data in two-dimensional Gaussian windows at multiple scales. These discrepancies can be taken advantage to effectively reduce the risk of cycle skipping. Then, we propose to use ATpV performs structural constraints on the velocity gradient in the vertical and horizontal directions to suppress artifacts in the gradient and preserve the shap boundaries between geological layers \cite[]{esser2018total, songEfficientWavefieldInversion2020}. Finally, AD is applied to optimize the proposed objective function to ensure an efficient implementation \cite[]{richardson2018seismic, zhuGeneralApproachSeismic2021, song2023weighted}. Synthetic and field data tests demonstrate that the proposed MS{\_}ATpV-FWI is able to provide high-resolution, high-precision velocity inversion results, even in the condition of low-frequency components absence, poor initial models, and noise contamination.

\section{Methodology}

This paper introduces a novel FWI method, MS{\_}ATpV-FWI, based on M-SSIM, and ATpV regularization, facilitated by AD technique. In this section, we first provide an overview of the traditional FWI. Next, we explain the novelly proposed objective function based on M-SSIM and evaluate its effectiveness in reducing the risk of cycle skipping. Then, we present ATpV regularization and assess its capability in removing the artifacts of the velocity updates and preserving the sharp boundaries. Finally, we explain that we use AD technique to optimize the proposed MS{\_}ATpV-FWI.
\subsection{FWI}

We consider the constant density acoustic wave equation to simulate seismic waves, given by

\begin{equation}
\label{deqn_ex1a}
\dfrac{1}{v^2}\dfrac{\partial^2 u(\mathbf{x},t)}{\partial t^2} - \nabla^2 u(\mathbf{x},t)= s(\mathbf{x}_s,t),
\end{equation}
where $v$ represents the velocity model; $s$ indicates the source function; $u$ is the corresponding pressure wavefield; $\mathbf{x}$ and $t$ denote the indexes of spatial location and time, respectively; $\mathbf{x}_s$ means the source location; the operator $\nabla^2$ represents the Laplacian operator.

By using the receivers to collect the signals from the simulated wavefield $u(s,\mathbf{x},t)$ for the source $s$, predicted seismic data $d_{pre}(s,r,t,v)$ can be obtained, the formulation is as follows

\begin{equation}
\label{deqn_ex1a}
d_{pre}(s,r,t)= \int _\Omega u(s,\mathbf{x},t) \cdot r(\mathbf{x}_r)d\mathbf{x},
\end{equation}
where $r$ and $\Omega$ represent the weight function of the receivers and the spatial range of the reception, respectively. $\mathbf{x}_r$ is the receivers location. The goal of FWI is to retrieve the velocity model $v$ by minimizing the misfit between predicted seismic data $d_{pre}(s,r,t,v)$ and observed seismic data $d_{obs}(s,r,t)$.

Traditional FWI utilizes the $l_2$ norm to construct the objective function, and the expression is given as follows

\begin{equation}
\label{deqn_ex1a}
J_{FWI}(v) = \dfrac{1}{2} \sum_{s} \sum_{r}\int_{t} \parallel d_{pre}(s,r,t,v) - d_{obs}(s,r,t)\parallel_{2}^{2} dt,
\end{equation}

The above objective function can be optimized by gradient descent method \cite[]{tarantolaInversionSeismicReflection1984}, and its gradient for velocity parameter $v$ is expressed as follows

\begin{equation}
\label{deqn_ex1a}
\dfrac{\partial J_{FWI}(v)}{\partial v} = \sum_{s} \sum_{r}\int_{t} [\dfrac{\partial d_{pre}(s,r,t,v)} {\partial v}b_{adj}(s,r,t,v)] dt,
\end{equation}
where $b_{adj}(s,r,t,v)$ represents the adjoint source. In the $l_2$ norm objective function, the adjoint source is the residual between the predicted and observed seismic data, which is given by $b_{adj}(s,r,t,v) = d_{pre}(s,r,t,v) - d_{obs}(s,r,t)$. Traditional FWI primarily focuses on a direct measurement of the amplitude difference between predicted and observed seismic data. If the seismic data lack low-frequency components or the initial velocity model is poor, the risk of cycle skipping is high.

\subsection{M-SSIM Objective Function}

To mitigate the risk of cycle skipping, we consider the differences between predicted data and observed data in both time and space directions across multi-scale structural features, and construct an objective function based on multi-scale structural similarity index measure (M-SSIM) as follows

\begin{equation}
\label{deqn_ex1a}
J_{MS}(v) = - \sum_{s} \sum_{r}\int_{t} MS(d_{pre}(s,r,t,v) , d_{obs}(s,r,t)) dt,
\end{equation}
where the symbol $MS$ represents the M-SSIM operation. M-SSIM compares the mean, variance, and covariance of predicted and observed data within Gaussian window with different scales in both time and space directions, as described by the following formulation

\begin{equation}
\label{deqn_ex1a}
MS(d_{pre} , d_{obs}) = \sum_{i=1}^n L (d_{pre(i)} , d_{obs(i)}) \cdot C (d_{pre(i)} , d_{obs(i)}) \cdot S (d_{pre(i)} , d_{obs(i)}),
\end{equation}
where $L$, $C$, and $S$ represent the mean contrast factor, variance contrast factor, and covariance contrast factor, respectively. The index $i$ denotes the $i$-th scale.

The mean contrast factor $L$ can measure the amplitude differences between the predicted and observed data in both time and space directions. Assuming that at the the $i$-th Gaussian window scale, the mathematical operation for $L$ is as follows

\begin{equation}
\label{deqn_ex1a}
L(d_{pre(i)} , d_{obs(i)}) = \dfrac{2 \mu_{d_{pre(i)}}\mu_{d_{obs(i)}}+C_1}{\mu_{d_{pre(i)}}^2 + \mu_{d_{obs(i)}}^2 +C_1},
\end{equation}
where $\mu$ represents the mean value of the data within the Gaussian window and $C_1$ is a constant. When the amplitude difference between the predicted and observed seismic data is small, $L$ approaches 1, indicating a high similarity between the predicted and observed data. When the difference is large, $L$ approaches -1, indicating a significant misfit.

Similarly, the variance contrast factor $C$ can quantify the difference of amplitude changes in both time and space directions. The mathematical operation for $C$ is as follows

\begin{equation}
\label{deqn_ex1a}
C(d_{pre(i)} , d_{obs(i)}) = \dfrac{2 \sigma_{pre(i)}\sigma_{obs(i)} + C_2}{\sigma_{d_{pre(i)}}^2 + \sigma_{d_{obs(i)}}^2 +C_2},
\end{equation}
where $\sigma$ indicates the standard deviation value of the data within the Gaussian window and $C_2$ is a constant. When $C$ is close to 1, it indicates that the difference in amplitude changes between the predicted and observed data is small. If $C$ is close to -1, it means that the difference is large.

Finally, the covariance contrast factor $S$ can measure the phase differences in both time and space directions. The mathematical operation for $S$ is as follows

\begin{equation}
\label{deqn_ex1a}
S(d_{pre(i)} , d_{obs(i)}) = \dfrac{\sigma_{d_{pre(i)}d_{obs(i)}}+C_3}{\sigma_{pre(i)}\sigma_{obs(i)} + C_3},
\end{equation}
where $C_3$ is a constant. If the phase of the predicted data is similar to that of the observed data, $S$ approaches 1. Otherwise, $S$ approaches -1.

\begin{figure}
\noindent\includegraphics[width=\textwidth]{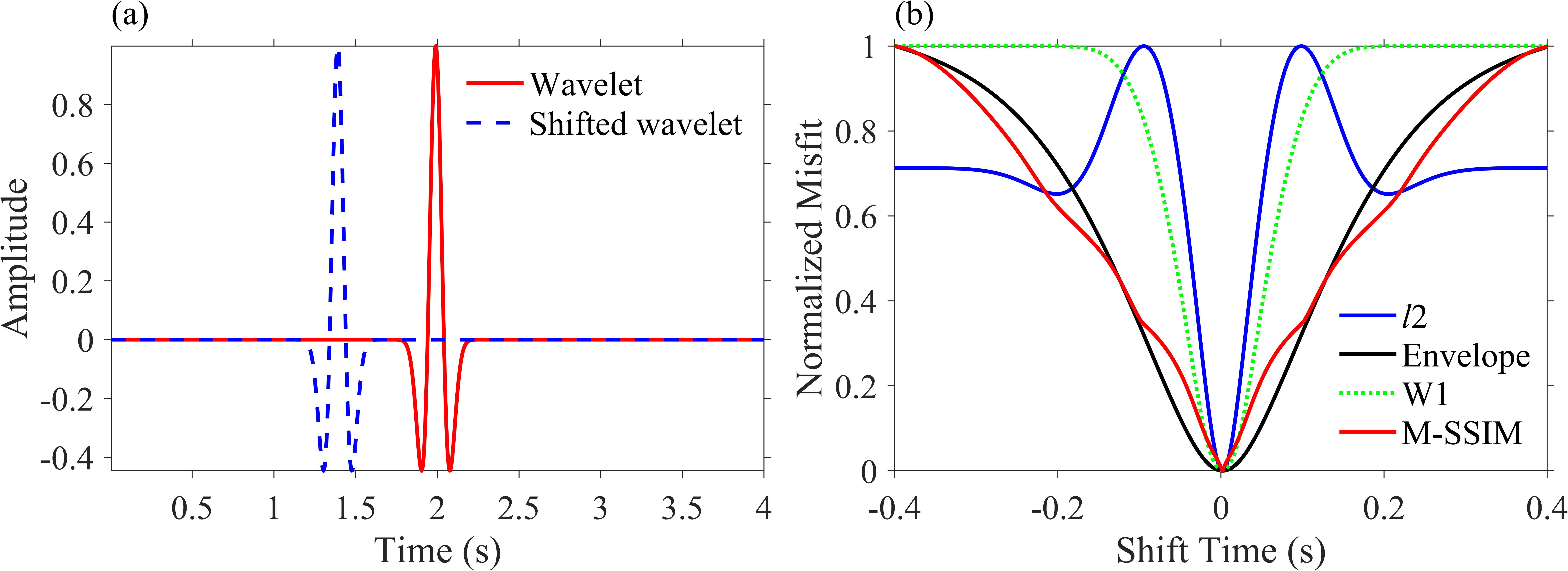}
\caption{Ricker wavelet misfit function test. (a) Ricker wavelet. (b) Nomalization misfit.}
\label{./misfit.jpg}
\end{figure}

The combination of these differences characterizes the multi-scale structural differences between preticted and observed seismic data. To fully extract the multi-scale structural features in seismic data and control the computational efficiency, the Gaussian window size in M-SSIM is set to $\lambda/4$, $\lambda/6$, and $\lambda/8$, where $\lambda$ represents the seismic wavelength. Using a Ricker wavelet and a shifted Ricker wavelet, we evaluate the performance of the proposed objective function, compared to $l_2$ norm, envelope, and W1 objective functions. The results, shown in Figure \ref{./misfit.jpg}, demonstrate that the $l_2$ function falls into local minima due to cycle skipping, while the envelope, W1, and the proposed misfit function effectively reduce the cycle skipping risk and converge to the global minimum. However, the objective function of envelope inversion converges slower when the time shifts are relatively small, resulting in lower resolution. While the objective function of W1-based FWI does not converge when the time shifts are relatively large. In contrast, the proposed M-SSIM objective function exhibits a reasonably good convergence in the whole optimization process.

\begin{figure}
\centering
\noindent\includegraphics[width=2in]{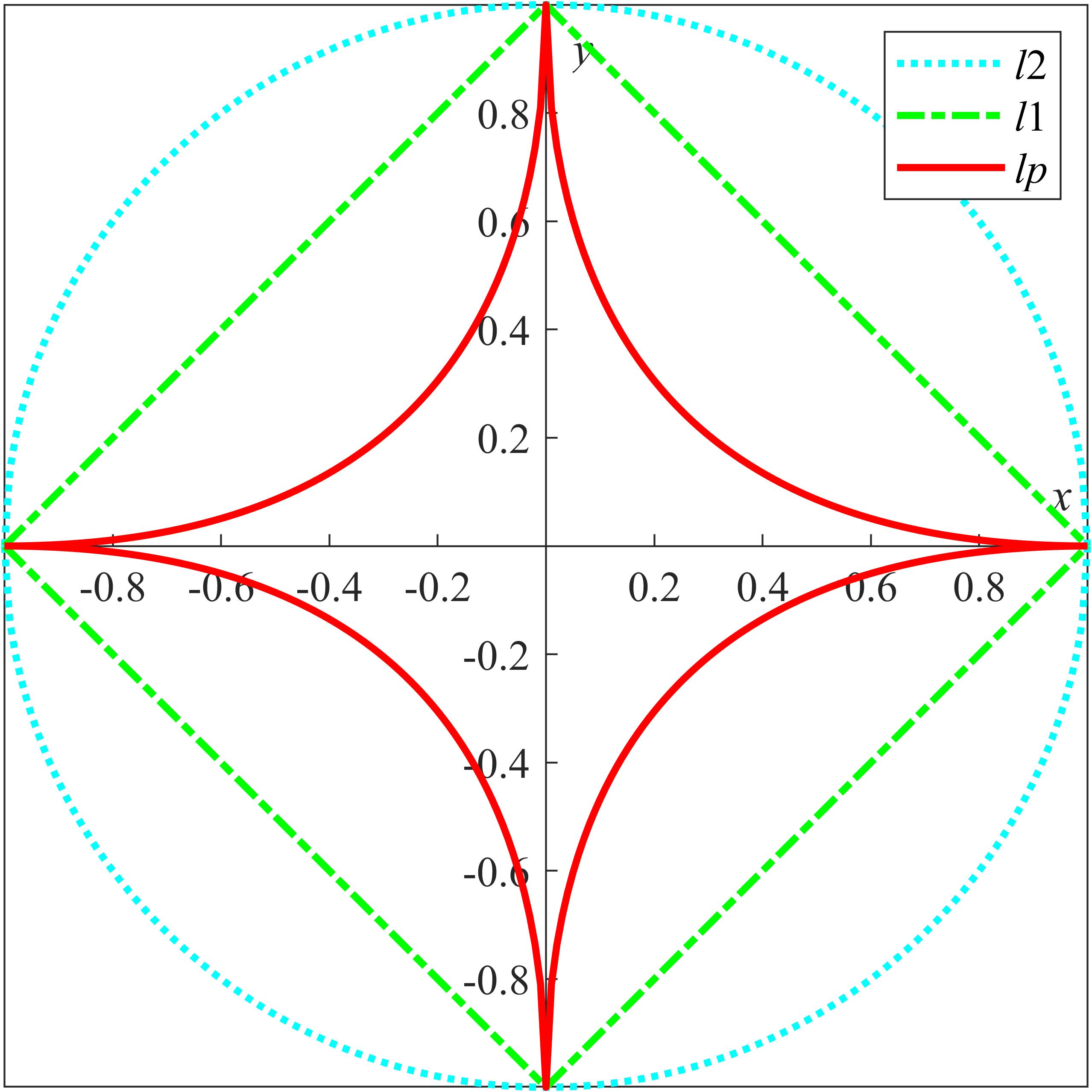}
\caption{Norm Ball.}
\label{norm.jpg}
\end{figure}

\subsection{ATpV}

Anisotropic total variation (ATV) regularization can impose structural constraints on velocity gradients, enhancing the inversion's ability to resolve underground structure boundaries. The formulation of the ATV regularization is as follows

\begin{align}
\parallel v \parallel_{ATV} = \| vD_x \|_1+\| D_zv \|_1,
\end{align}
where $D_x$ and $D_y$ indicate the horizontal and vertical difference matrices respectively, and the formulations are expressed as follows 

\begin{equation}
\renewcommand{\arraystretch}{0.5} 
D_x = \begin{bmatrix}
-1 & 0 & \cdots & 0\\
1 & -1 & \ddots & \vdots\\
0 & 1 & \ddots & 0 \\
\vdots & \ddots & \ddots & -1 \\
0 & \cdots & 0 & 1
\end{bmatrix}_{x \times (x-1)}
\end{equation}

\begin{equation}
\renewcommand{\arraystretch}{0.5} 
D_z = \begin{bmatrix}
-1 & 1 & 0 & \cdots & 0\\
0 & -1 & 1 & \ddots & \vdots\\
\vdots & \ddots & \ddots & \ddots & 0\\
0 & \cdots & 0 & -1 & 1
\end{bmatrix}_{(z-1) \times z}
\end{equation}
where $x$ and $z$ represent the horizontal and vertical spatial grid sizes of the model $v$, respectively.

However, traditional ATV regularization uses $l_1$ norm for structural constraints, which has low constraint ability and cannot fully suppress artifacts in the gradient. To address this, we employ the $l_p$ quasi-norm to construct ATpV regularization \cite[]{he2022seismic}, which can more effectively suppress gradient artifacts and preserve the shap boundaries between geological layers. The formulations of the ATpV regularization is expressed as follows

\begin{align}
\parallel v \parallel_{ATpV} = \| vD_x \|_p^p + \| D_yv \|_p^p,
\end{align}

Figure \ref{norm.jpg} illustrates the norm balls for the $l_2$ norm, $l_1$ norm, and $l_p$ quasi-norm. As shown, the solution space of the $l_p$ quasi-norm is closer to the coordinate axes, indicating that it better suppresses artifacts in the gradient.

\subsection{MS{\_}ATpV-FWI}
\begin{figure}
\centering
\noindent\includegraphics[width=5in]{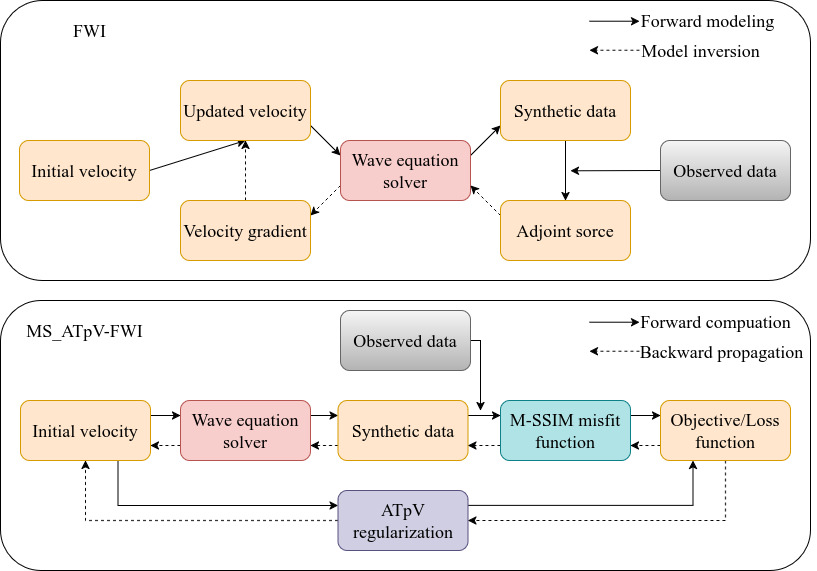}
\caption{Workflow of MS{\_}ATpV-FWI and FWI.}
\label{workflows.jpg}
\end{figure}

By combining the M-SSIM misfit function with ATpV regularization, the objective function is constructed as shown below

\begin{equation}
\label{obj}
J_{}(v) = J_{MS}(v) + \lambda \cdot \parallel v \parallel_{ATpV},
\end{equation}
where $\lambda$ represents the regularization weight parameter.

For the objective function shown in equation (\ref{obj}), solving the adjoint source is challenging. To address this, we use automatic differentiation (AD) to solve the objective function directly \cite[]{zhuGeneralApproachSeismic2021, richardson2022deepwave}. AD is a key technology in deep learning. It can efficiently and automatically calculate the gradient of functions and reduce the complexity of manual calculations. The principle of AD is to convert complex objective functions into calculation graphs and then apply the chain rule to trace through these graphs, obtaining precise gradients for each parameter. Unlike finite difference and symbolic differentiation, AD avoids truncation and rounding errors, providing flexible, efficient calculations independent of input dimension.

In AD-based FWI, the objective function acts as the loss function in deep learning, and the velocity model to be inverted is treated as the trainable network parameter. The workflow of the proposed MS{\_}ATpV-FWI method using AD and the traditional FWI using adjoint-state method are illustrated in Figure \ref{workflows.jpg}. Traditional FWI requires the derivation of the adjoint source and then the manual calculation of the gradient information. The proposed method MS{\_}ATpV-FWI can automatically calculate the velocity gradient by backward propagation the defined objection or loss function.

\begin{figure}
\centering
\noindent\includegraphics[width=3.1in]{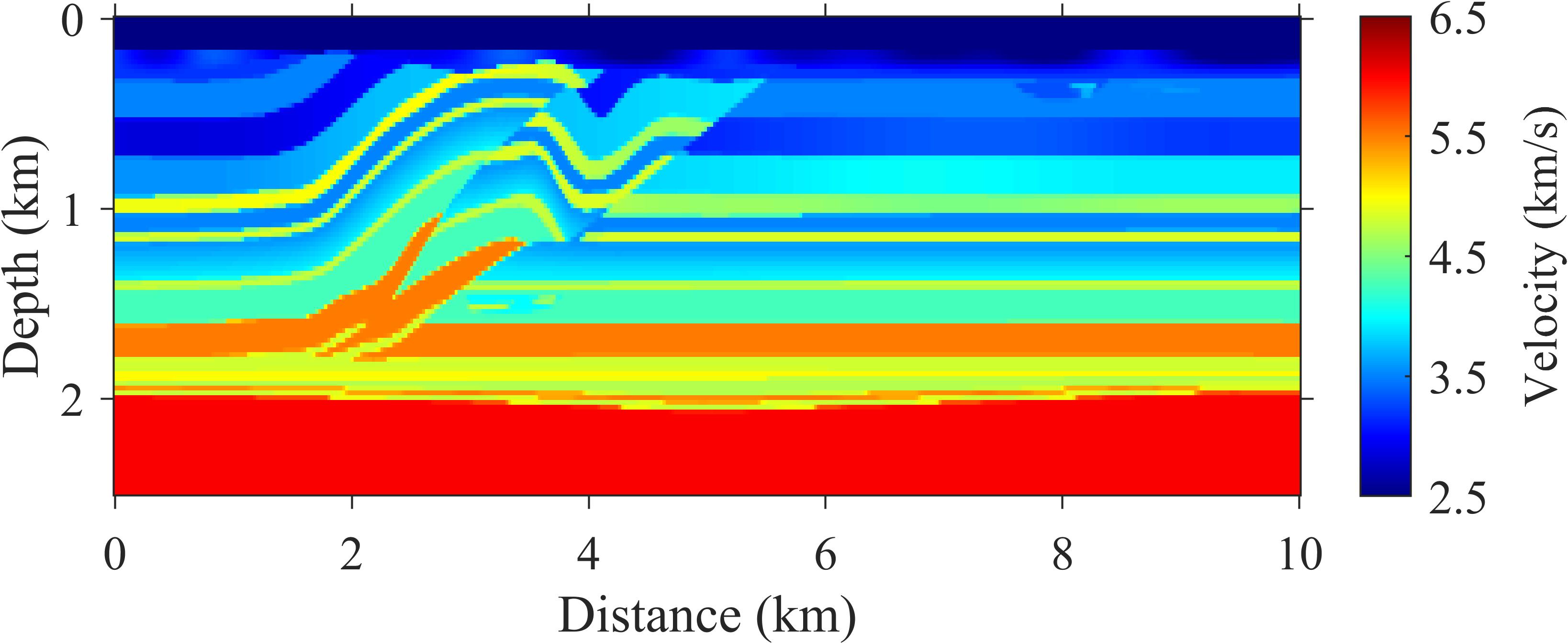}
\caption{Overthrust velocity model.}
\label{./overthrust/overthrustTI.jpg}
\end{figure}

\begin{figure}
\centering
\noindent\includegraphics[width=\textwidth]{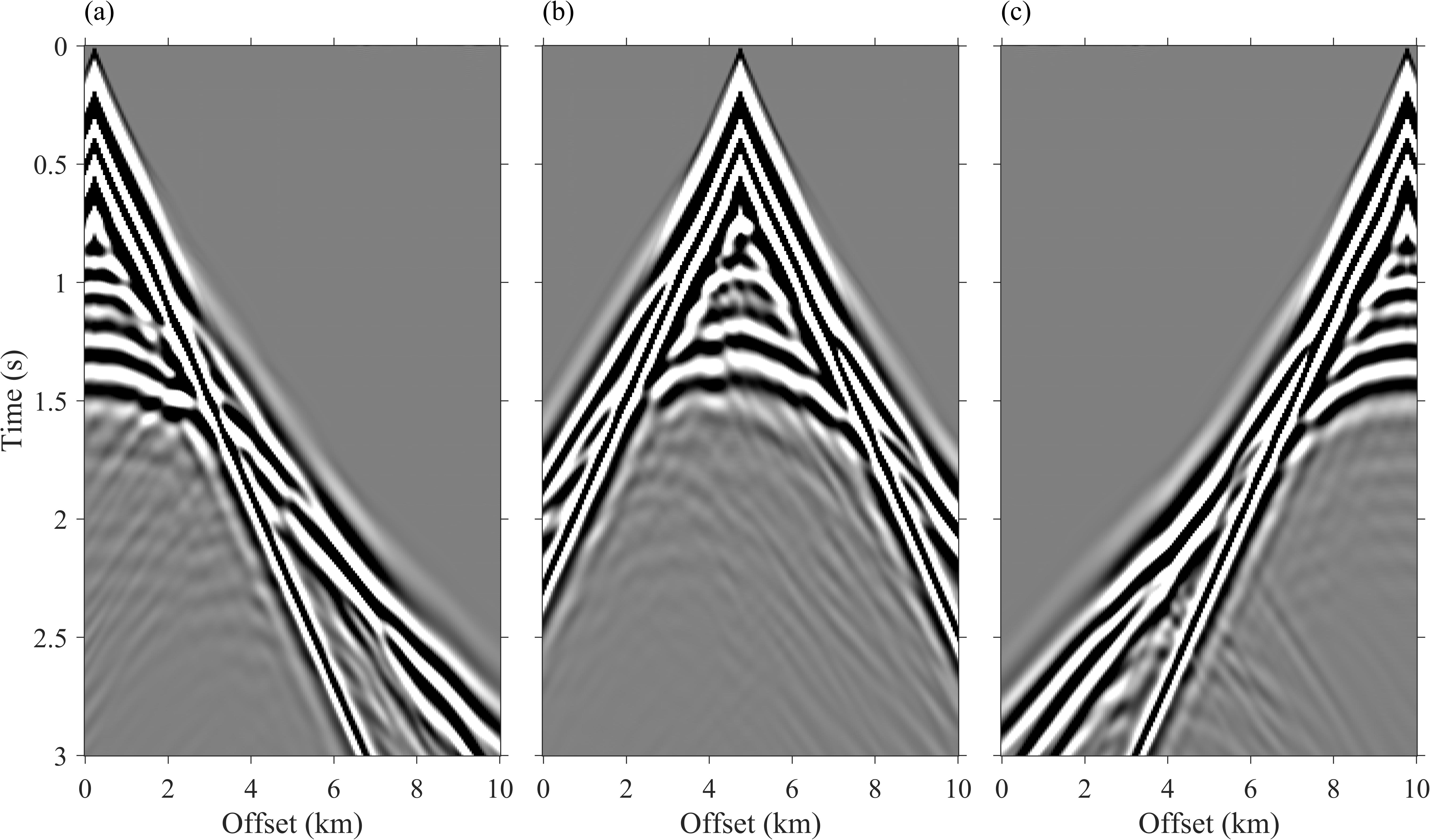}
\caption{Overthrust synthetic data. (a) The 1st shot. (b) The 10th shot. (c) The 20th shot.}
\label{./overthrust/d_obs.jpg}
\end{figure}

\begin{figure}
\centering
\noindent\includegraphics[width=\textwidth]{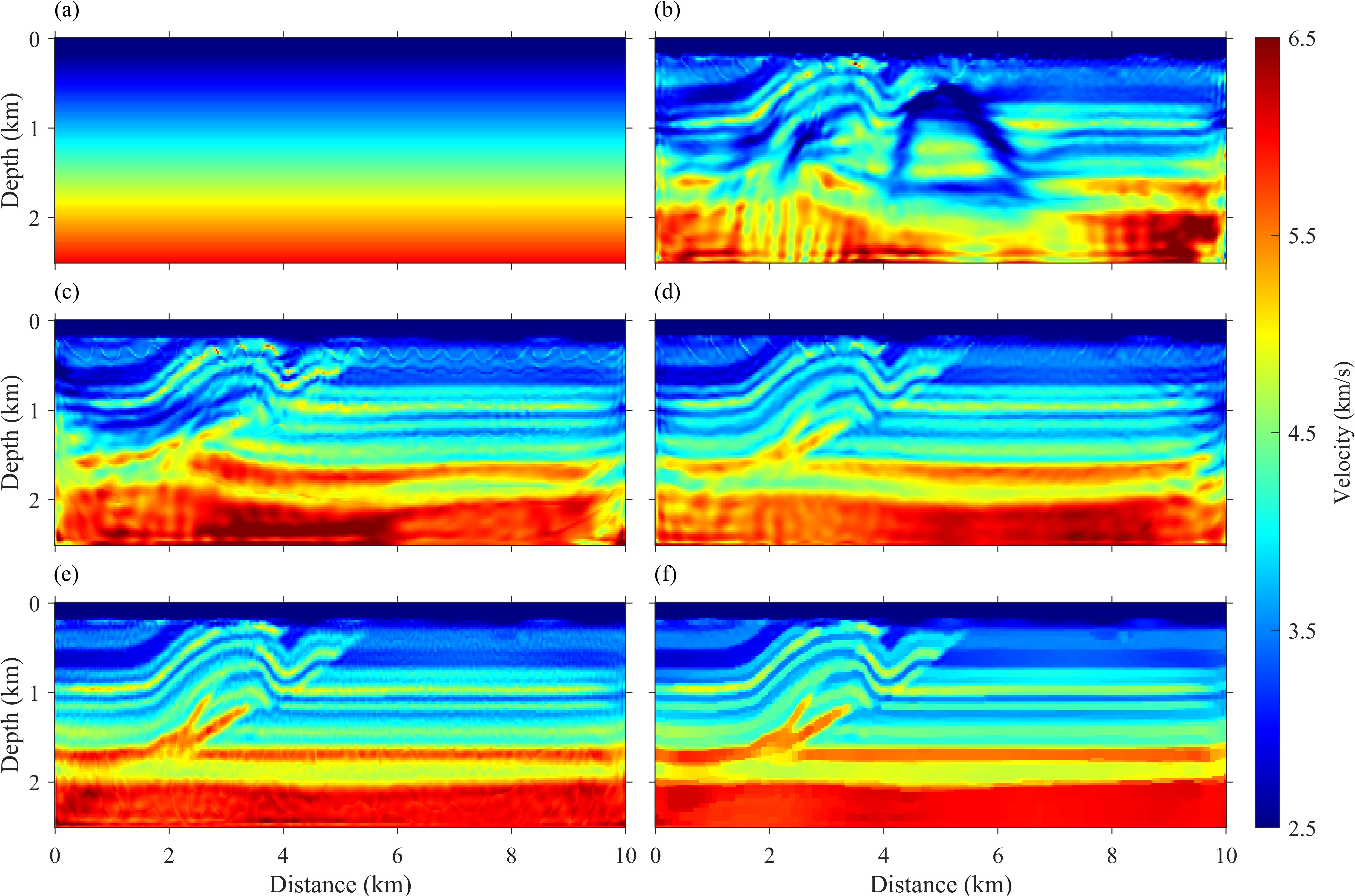}
\caption{Overthrust model inversion results. (a) Initial model. (b) FWI. (c) EI. (d) W1-FWI. (e) MS-FWI. (f) MS{\_}ATpV-FWI.}
\label{./overthrust/overthrust.jpg}
\end{figure}

\begin{figure}
\centering
\noindent\includegraphics[width=\textwidth]{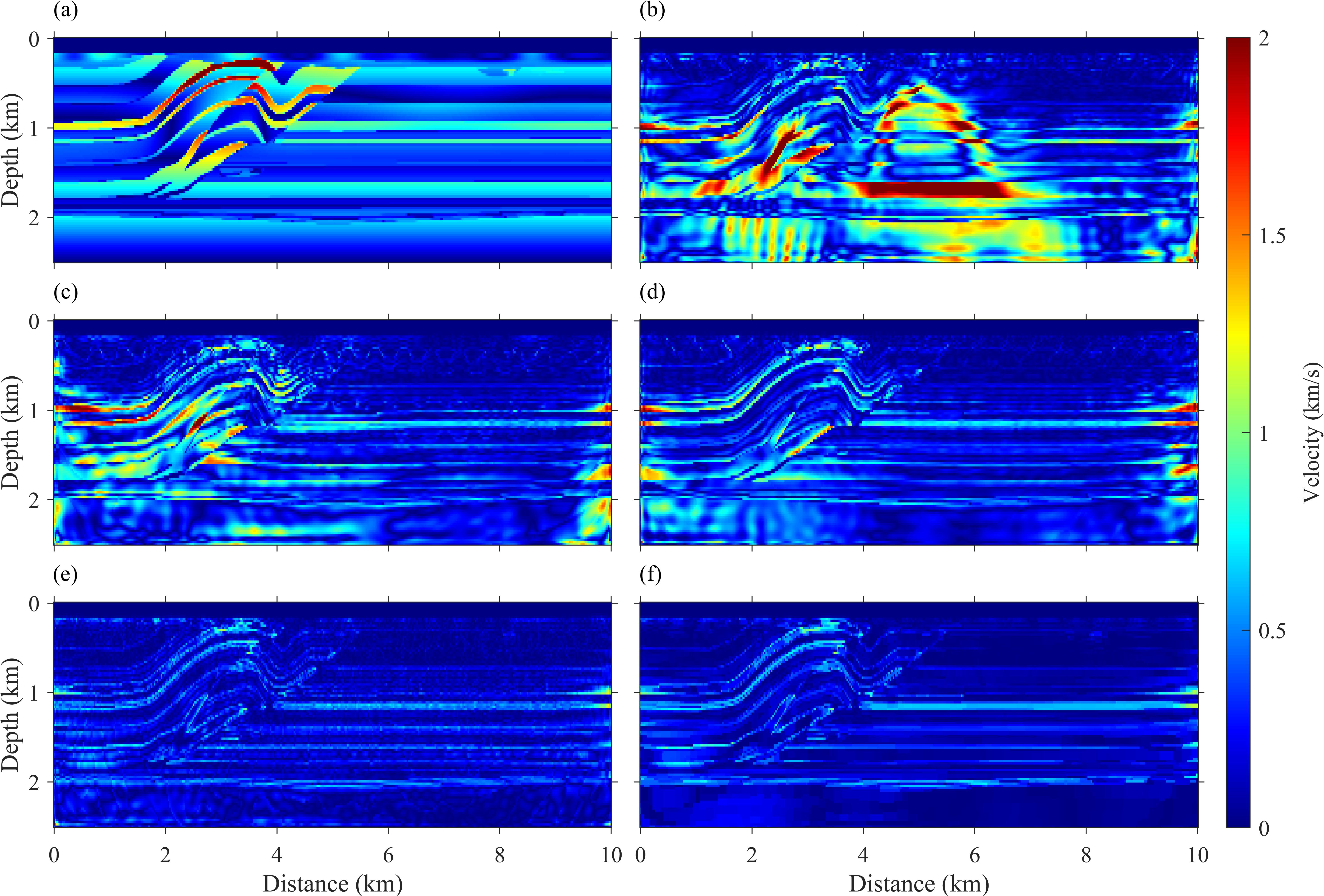}
\caption{Overthrust model absolute errors. (a) Initial model. (b) FWI. (c) EI. (d) W1-FWI. (e) MS-FWI. (f) MS{\_}ATpV-FWI.}
\label{./overthrust/overthrustr.jpg}
\end{figure}

\section{Experiments and Results}

In this section, we compare the performance of the proposed MS{\_}ATpV-FWI method with different objective functions using three synthetic datasets and one field seismic dataset. For the synthetic data tests, we use seismic data without low-frequency information below 3 Hz contaminated by random noise, with a poor initial model, and add random noise. The specific experimental setup is as follows.

\subsection{Overthrust Model}

First, we test the feasibility of the proposed MS{\_}ATpV-FWI method using the overthrust model \cite[]{aminzadeh1994seg}. The model has a spatial grid size of $100\times400$ with a 25 m spatial sampling interval. The true velocity model is shown in Figure \ref{./overthrust/overthrustTI.jpg}. Using a Ricker wavelet with a main frequency of 6 Hz, we place 20 equally spaced seismic sources and arrange 200 receivers at equal intervals. The acoustic wave equation is applied to generate 3s synthetic seismic data. Frequency components below 3 Hz are filtered out, with the resulting synthetic data for three shot gathers are shown in Figures \ref{./overthrust/d_obs.jpg}a-\ref{./overthrust/d_obs.jpg}c.

Figure \ref{./overthrust/overthrust.jpg}a displays the initial velocity for the tests, which is a constant gradient model without details of the true model. Figures \ref{./overthrust/overthrust.jpg}b-\ref{./overthrust/overthrust.jpg}f shows the inversion results of FWI, EI, W1-FWI, MS-FWI, and MS{\_}ATpV-FWI, respectively. The inverted result of the conventional FWI exhibit inaccurate velocity structure due to cycle skipping. EI, focusing on envelope information, also struggles to capture complex velocity structures in the left part. W1-FWI improves the recovery of the structures in the shallow part, but it fails to construct the layers in the deep regions. In contrast, MS-FWI and MS{\_}ATpV-FWI account for multi-scale structural similarity between predicted and observed data, reducing the risk of cycle skipping, resulting in well-inverted velocity models shown in Figures \ref{./overthrust/overthrust.jpg}e and \ref{./overthrust/overthrust.jpg}f. While the introduction of ATpV regularization suppresses the velocity gradient artifacts and preserves the sharp boundaries of the layer interfaces. Figures \ref{./overthrust/overthrustr.jpg}a-\ref{./overthrust/overthrustr.jpg}f presents the absolute errors between the initial, inverted velocities and the true one. Among them, MS{\_}ATpV-FWI exhibits the smallest absolute error, further demonstrating the feasibility of the proposed method.

The root mean square error (RMSE) and structural similarity index measure (SSIM) are used to quantitatively evaluate the inversion results, and the results are shown in Table \ref{RS1}. The performance of the proposed method MS{\_}ATpV-FWI is significantly better than other methods.

\begin{table}
\caption{RMSE and SSIM of the Overthrust model inversion results.}
\label{RS1}
    \centering
    \begin{tabularx}{\textwidth}{c>{\centering\arraybackslash}X>{\centering\arraybackslash}X>{\centering\arraybackslash}X>{\centering\arraybackslash}Xc}
    \hline
    \multirow{1}{*}{{Method}} & \multicolumn{1}{c}{FWI} & \multicolumn{1}{c}{EI} & \multicolumn{1}{c}{W1-FWI} & \multicolumn{1}{c}{MS-FWI} & \multicolumn{1}{c}{MS{\_}ATpV-FWI}\\
    \hline
    RMSE & 645.54 & 406.07 & 332.53 & 212.94 & \bf{201.51} \\
    SSIM & 0.46 & 0.68 & 0.77 & 0.89 & \bf{0.91} \\
    \hline
    \end{tabularx}
\end{table}

\begin{figure}
\centering
\noindent\includegraphics[width=3.1in]{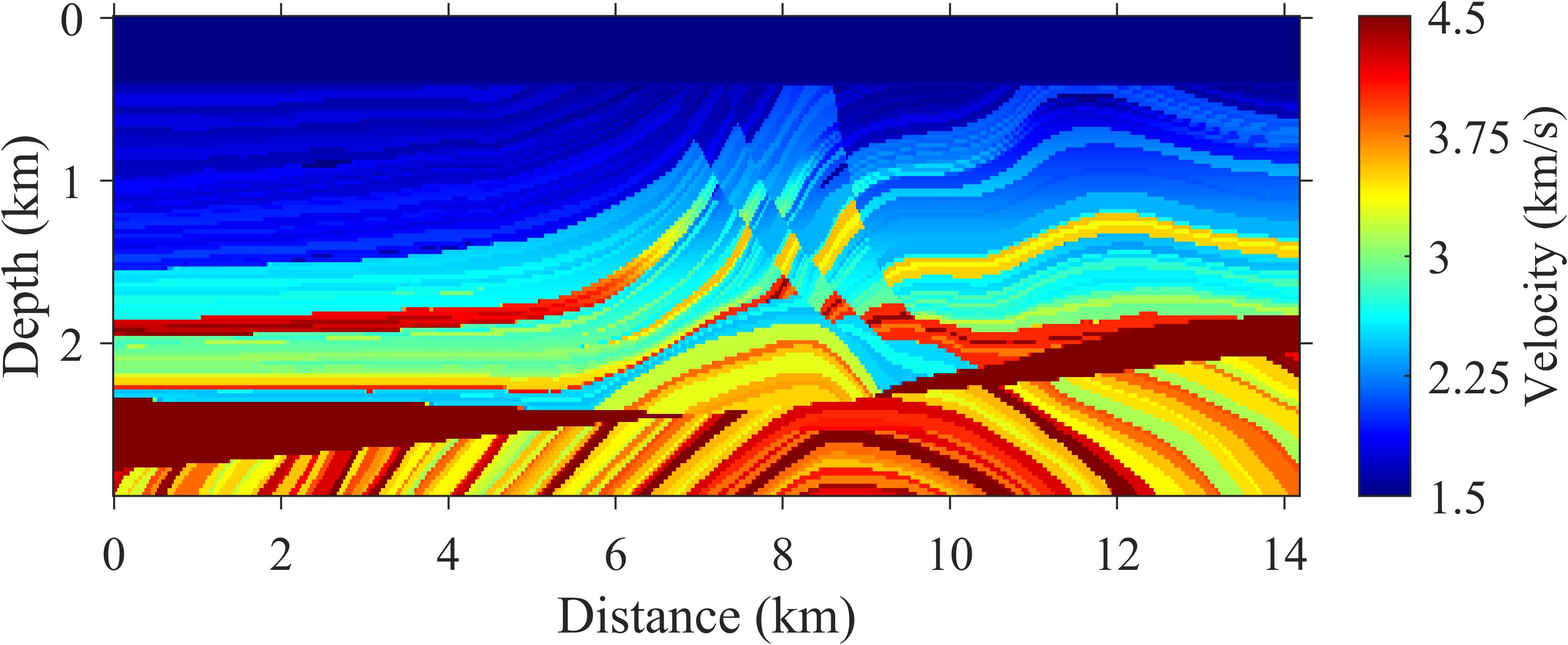}
\caption{Marmousi2 velocity model.}
\label{./marmousi/marmousiTI.jpg}
\end{figure}

\subsection{Marmousi2 Model}

\begin{figure}
\centering
\noindent\includegraphics[width=\textwidth]{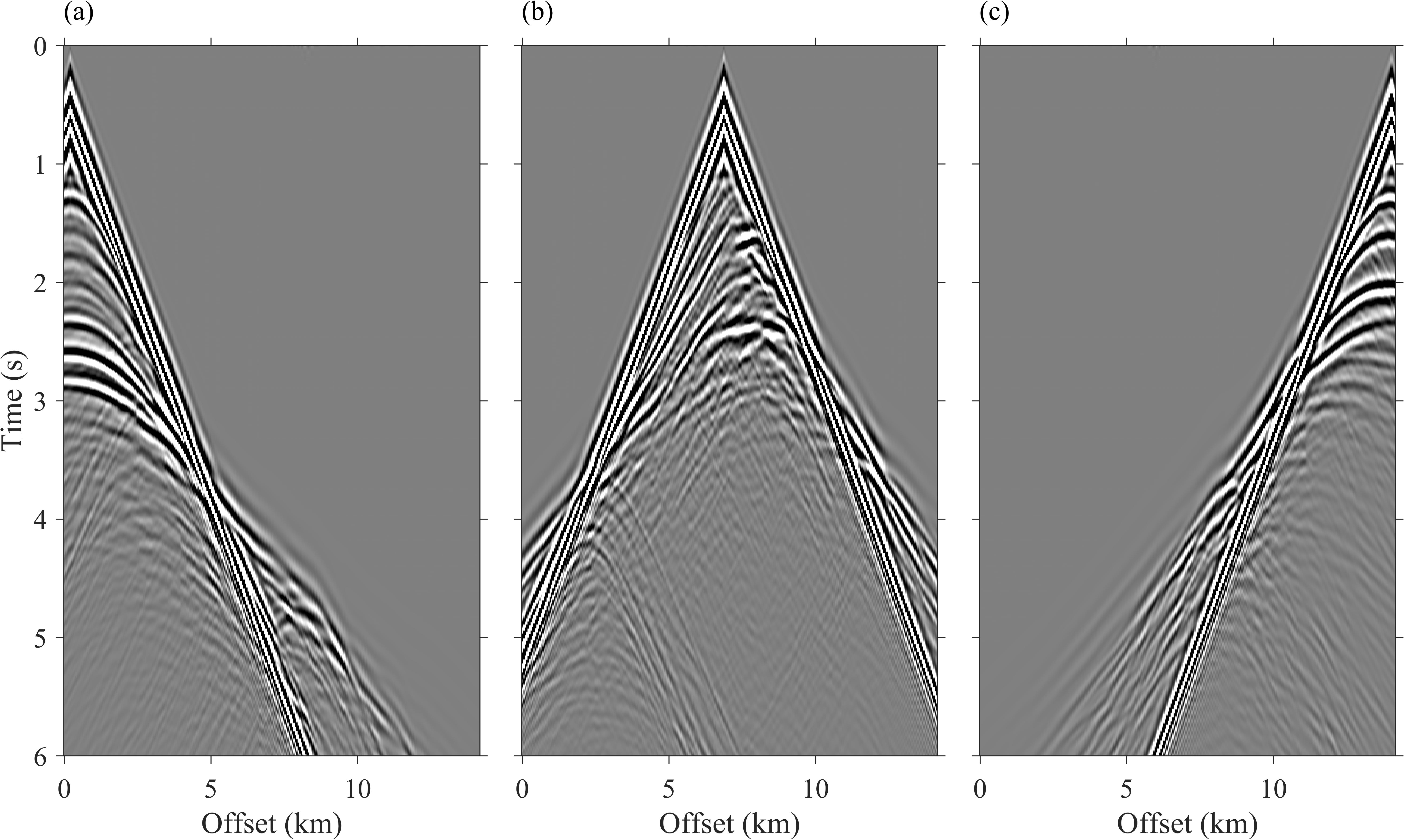}
\caption{Marmousi2 synthetic data. (a) The 1st shot. (b) The 15th shot. (c) The 30th shot.}
\label{./marmousi/d_obs.jpg}
\end{figure}

Next, we use the Marmousi2 model \cite[]{versteeg1994marmousi} to test the inversion performance of the proposed MS{\_}ATpV-FWI method for complex subsurface structures. The spatial grid size is $117\times567$, with a sampling interval of 30 m. The true velocity model is shown in Figure \ref{./marmousi/marmousiTI.jpg}. Using a Ricker wavelet with a main frequency of 6 Hz, 30 evenly spaced seismic sources and 284 evenly spaced receivers are deployed. The acoustic wave equation generates synthetic seismic data with record time of 6s, with frequency components below 3 Hz filtered out, and three shot gathers are shown in Figure \ref{./marmousi/d_obs.jpg}.

\begin{figure}
\centering
\noindent\includegraphics[width=\textwidth]{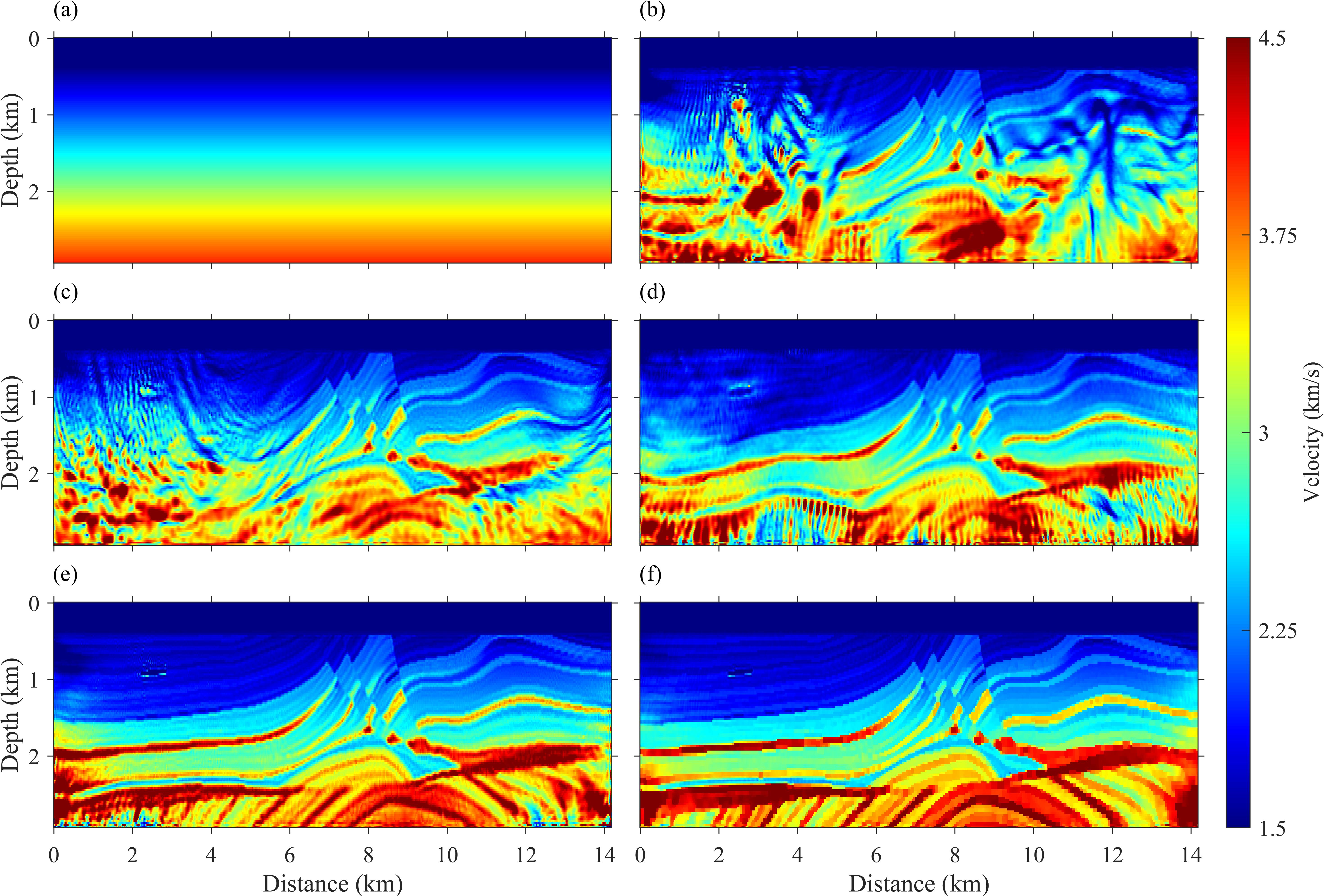}
\caption{Marmousi2 model inversion results. (a) Initial model. (b) FWI. (c) EI. (d) W1-FWI. (e) MS-FWI. (f) MS{\_}ATpV-FWI.}
\label{./marmousi/marmousi.jpg}
\end{figure}

\begin{figure}
\centering
\noindent\includegraphics[width=\textwidth]{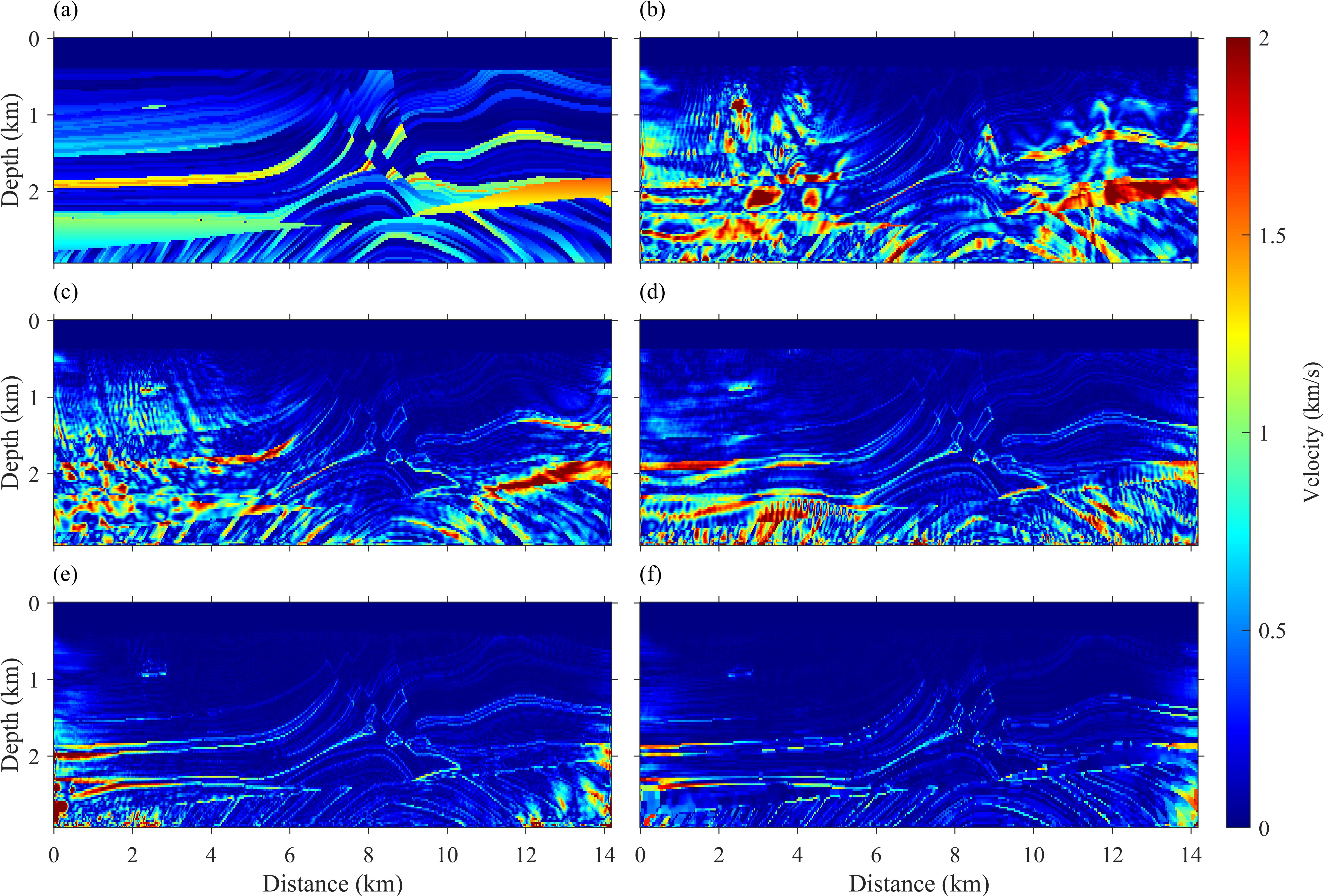}
\caption{Marmousi2 model absolute errors. (a) Initial model. (b) FWI. (c) EI. (d) W1-FWI. (e) MS-FWI. (f) MS{\_}ATpV-FWI.}
\label{./marmousi/marmousir.jpg}
\end{figure}

With a similar figure display setup as the previous example, Figure \ref{./marmousi/marmousi.jpg}a presents a constant-gradient initial velocity model. Figures \ref{./marmousi/marmousi.jpg}b-\ref{./marmousi/marmousi.jpg}f shows the inversion results of FWI, EI, W1-FWI, MS-FWI, and MS{\_}ATpV-FWI, respectively. For the Marmousi2 model, the inverted velocity from the conventional FWI exhibits significant cycle skipping artifacts. EI struggles to accurately recover velocities in thin layers as it cannot completely mitigate the cycle-skipping issue. W1-FWI manages to roughly recover the shallow structures, but it obtains inaccurate velocity structure in the deep regions. In contrast, MS-FWI and MS{\_}ATpV-FWI effectively capture the complex velocity structure. Moreover, ATpV regularization preserves the sharp boundaries of the layers and improves the continuity of the inversion result. Figures \ref{./marmousi/marmousir.jpg}a-\ref{./marmousi/marmousir.jpg}f displays the absolute errors between the initial, inverted velocities and the true one. The absolute errors for FWI, EI, W1-FWI, and MS-FWI are clearly larger than those for MS{\_}ATpV-FWI, demonstrating that the proposed method offers superior inversion performance, even for complex subsurface structures.

Table \ref{RS2} shows the RMSE and SSIM of the inversion results. For complex underground structures, the performance of the proposed method MS{\_}ATpV-FWI is still significantly better than other methods.

\begin{table}
\caption{RMSE and SSIM of the Marmousi2 model inversion results.}
\label{RS2}
    \centering
    \begin{tabularx}{\textwidth}{c>{\centering\arraybackslash}X>{\centering\arraybackslash}X>{\centering\arraybackslash}X>{\centering\arraybackslash}Xc}
    \hline
    \multirow{1}{*}{{Method}} & \multicolumn{1}{c}{FWI} & \multicolumn{1}{c}{EI} & \multicolumn{1}{c}{W1-FWI} & \multicolumn{1}{c}{MS-FWI} & \multicolumn{1}{c}{MS{\_}ATpV-FWI}\\
    \hline
    RMSE & 611.98 & 500.82 & 458.19 & 307.58 & \bf{224.34} \\
    SSIM & 0.39 & 0.53 & 0.68 & 0.86 & \bf{0.91} \\
    \hline
    \end{tabularx}
\end{table}

\begin{figure}
\centering
\noindent\includegraphics[width=3.1in]{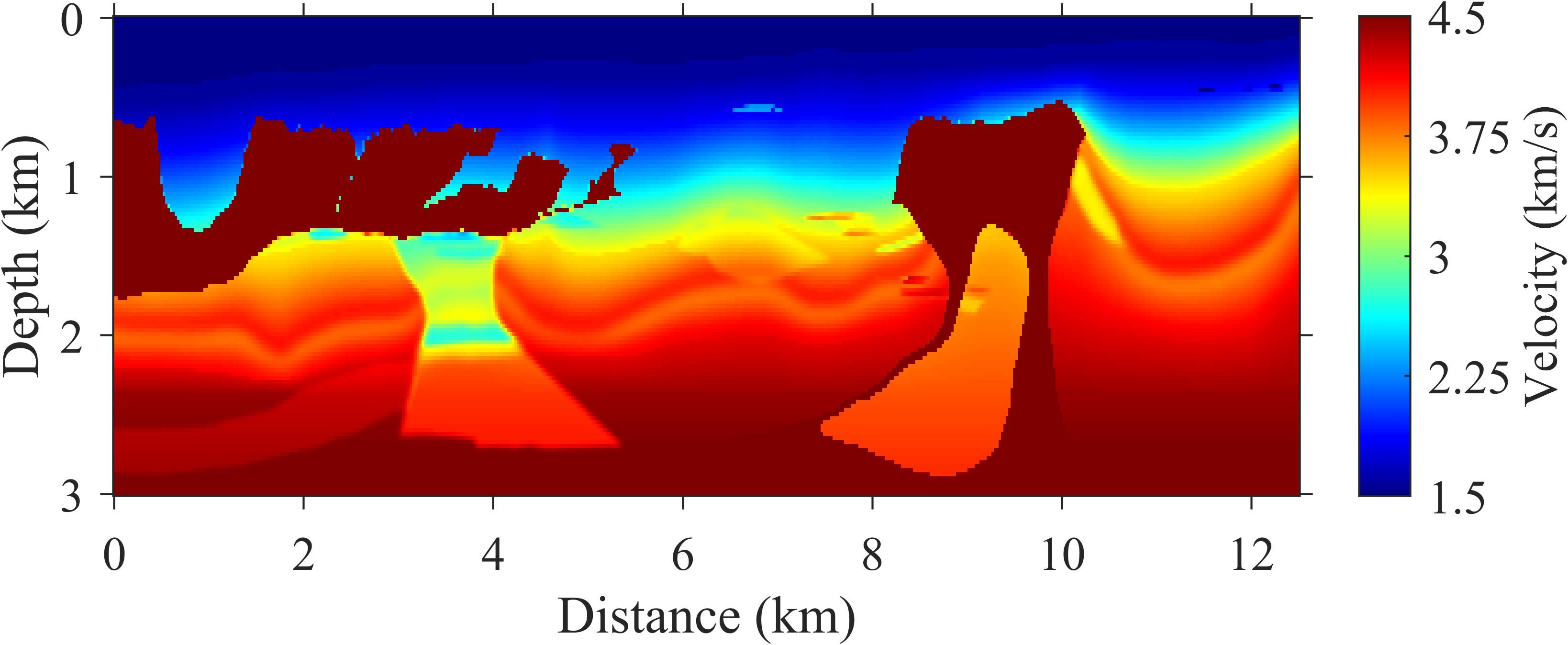}
\caption{2004 BP velocity model.}
\label{./BP/BPTI.jpg}
\end{figure}

\subsection{2004 BP Model}

\begin{figure}
\centering
\noindent\includegraphics[width=\textwidth]{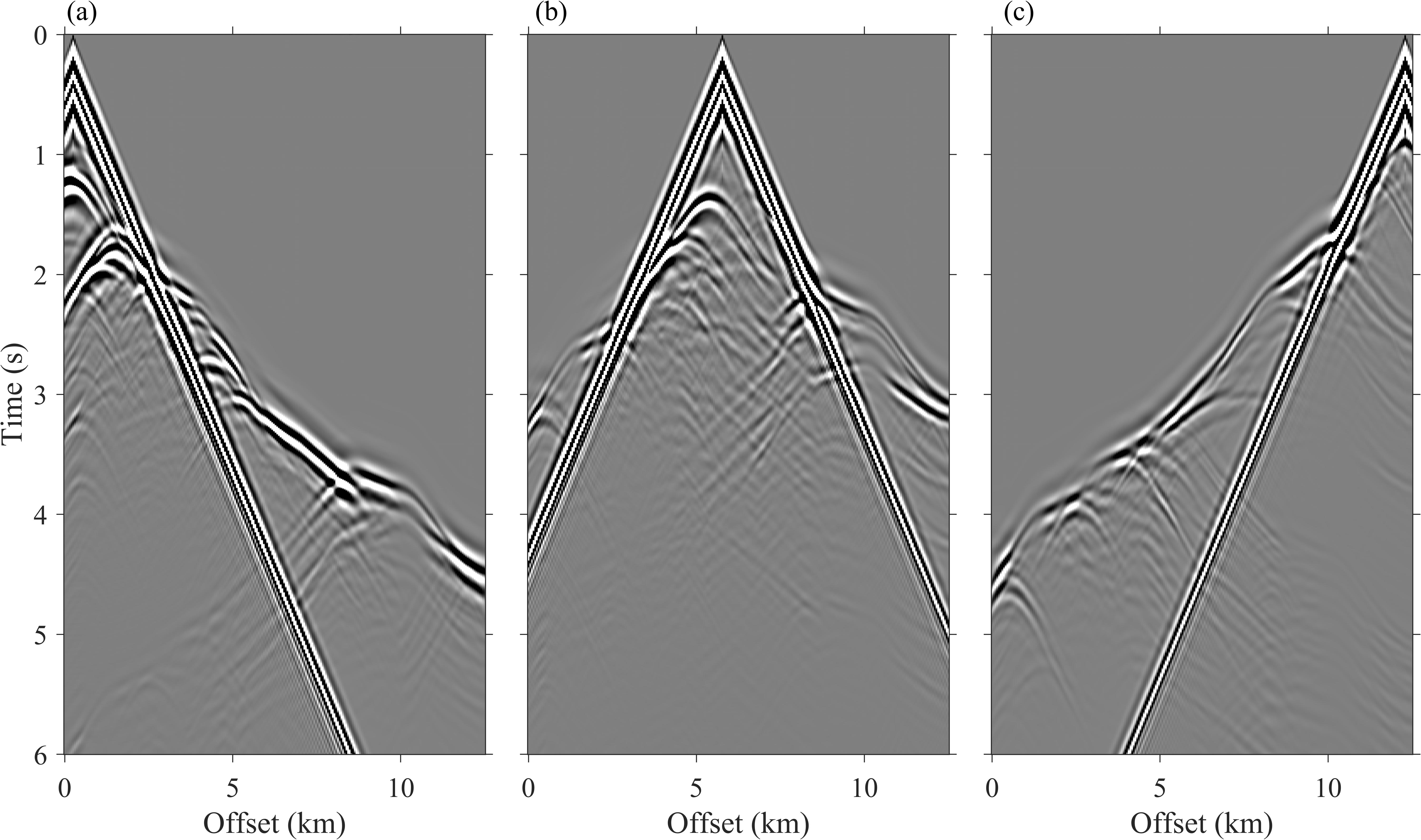}
\caption{2004 BP synthetic data. (a) The 1st shot. (b) The 12th shot. (c) The 25th shot.}
\label{./BP/d_obs.jpg}
\end{figure}

Then, we consider the 2004 BP model \cite[]{billette20052004} to test the inversion performance of the proposed MS{\_}ATpV-FWI method for salt body reconstruction. The spatial grid size is $120\times500$ with a sampling interval of 25m. The true velocity model is shown in Figure \ref{./BP/BPTI.jpg}. Using a Ricker wavelet with a main frequency of 6 Hz, 25 evenly spaced seismic sources and 250 equally spaced receivers are deployed. The acoustic wave equation generates 6s of synthetic seismic data, with frequency components below 3 Hz filtered out, and three shot gathers are shown in Figure \ref{./BP/d_obs.jpg}.

\begin{figure}
\centering
\noindent\includegraphics[width=\textwidth]{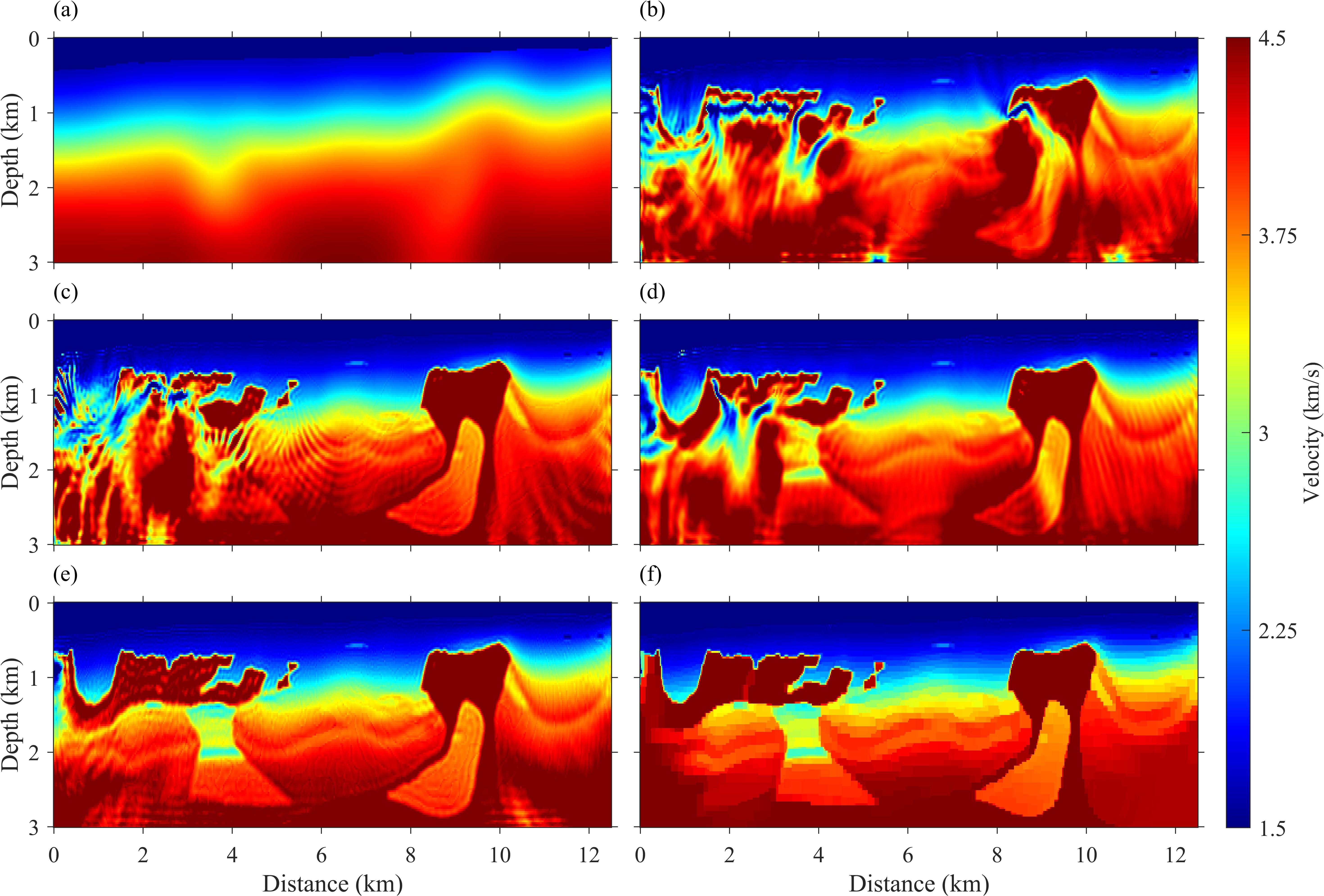}
\caption{2004 BP model inversion results. (a) Initial model. (b) FWI. (c) EI. (d) W1-FWI. (e) MS-FWI. (f) MS{\_}ATpV-FWI.}
\label{./BP/BP.jpg}
\end{figure}

\begin{figure}
\centering
\noindent\includegraphics[width=\textwidth]{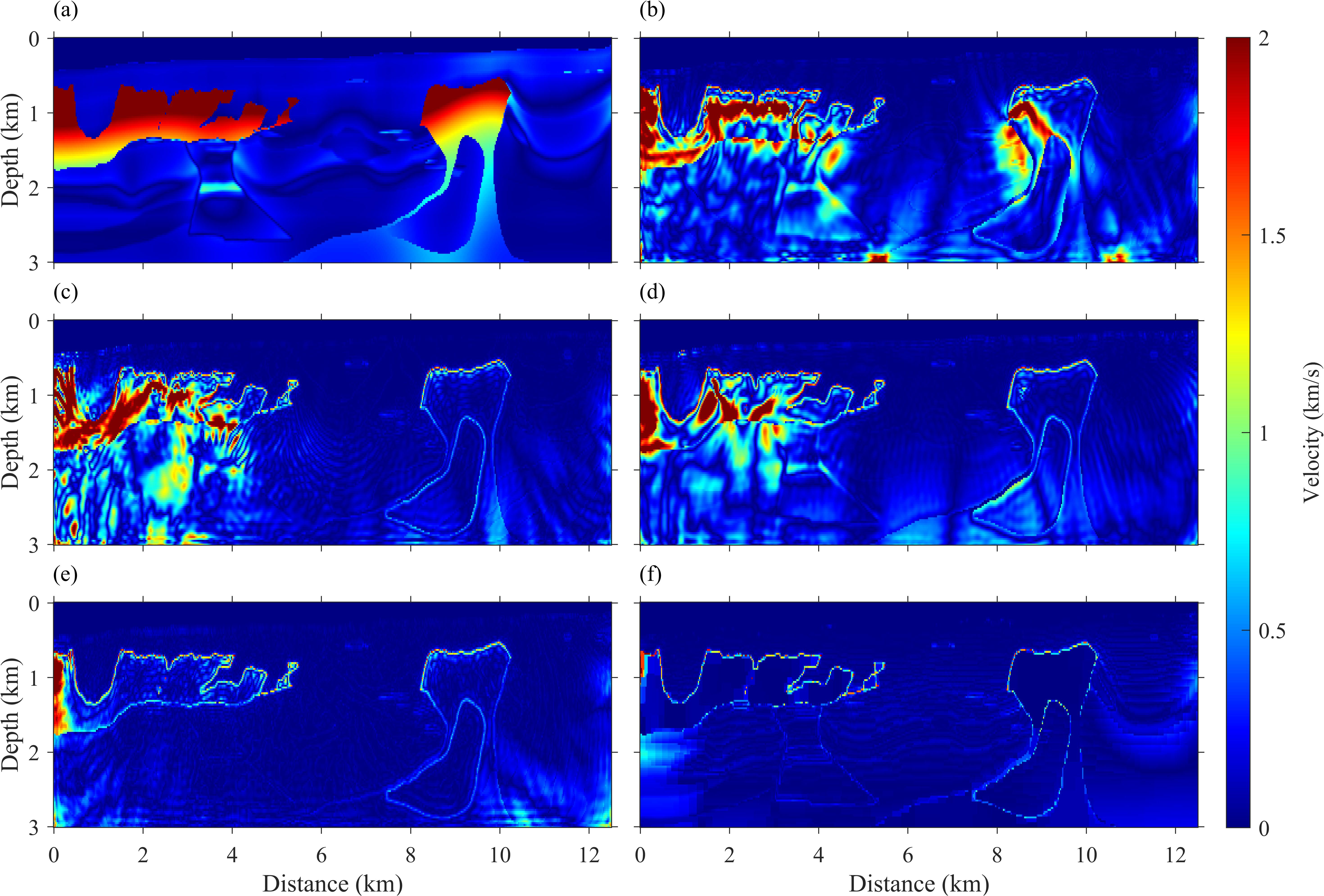}
\caption{2004 BP model absolute errors. (a) Initial model. (b) FWI. (c) EI. (d) W1-FWI. (e) MS-FWI. (f) MS{\_}ATpV-FWI.}
\label{./BP/BPr.jpg}
\end{figure}

In this example, we use a smooth background velocity without any salt body information as the initial velocity, as shown in Figure \ref{./BP/BP.jpg}a. After 600 number of iterations, we obtain the inversion results for FWI, EI, W1-FWI, MS-FWI, and MS{\_}ATpV-FWI in Figures \ref{./BP/BP.jpg}b-\ref{./BP/BP.jpg}f, respectively. It is evident that FWI falls into a local minima due to cycle skipping, failing to accurately describe the spatial morphology of the salt dome. The EI and W1-FWI methods struggle to capture the detailed spatial structure of the complex salt dome on the left. In contrast, MS-FWI and MS{\_}ATpV-FWI can successfully identify the locations of the two salt domes and accurately characterizes their structures, while the proposed MS{\_}ATpV-FWI method has a stronger ability to preserve sharp boundaries and improve continuity. Figure \ref{./BP/BPr.jpg}a-\ref{./BP/BPr.jpg}f shows the absolute errors between the initial, inverted velocities and the true one. It can be seen that FWI, EI, and W1-FWI exhibit large errors at the positions where are the salt bodies located. While MS{\_}ATpV-FWI demonstrates the smallest absolute error among them, further confirming that the proposed method can effectively characterize complex subsurface structural features, including salt domes.

Table \ref{RS3} shows the RMSE and SSIM of the inversion results for different FWI strategies. For characterizing the subsurface salt dome structure, the performance of the proposed method MS{\_}ATpV-FWI is still significantly better than other methods.

\begin{table}
\caption{RMSE and SSIM of the 2004 BP model inversion results.}
\label{RS3}
    \centering
    \begin{tabularx}{\textwidth}{c>{\centering\arraybackslash}X>{\centering\arraybackslash}X>{\centering\arraybackslash}X>{\centering\arraybackslash}Xc}
    \hline
    \multirow{1}{*}{{Method}} & \multicolumn{1}{c}{FWI} & \multicolumn{1}{c}{EI} & \multicolumn{1}{c}{W1-FWI} & \multicolumn{1}{c}{MS-FWI} & \multicolumn{1}{c}{MS{\_}ATpV-FWI}\\
    \hline
    RMSE & 488.60 & 388.19 & 346.76 & 190.66 & \bf{138.02} \\
    SSIM & 0.59 & 0.68 & 0.76 & 0.90 & \bf{0.92} \\
    \hline
    \end{tabularx}
\end{table}

\begin{figure}
\centering
\noindent\includegraphics[width=\textwidth]{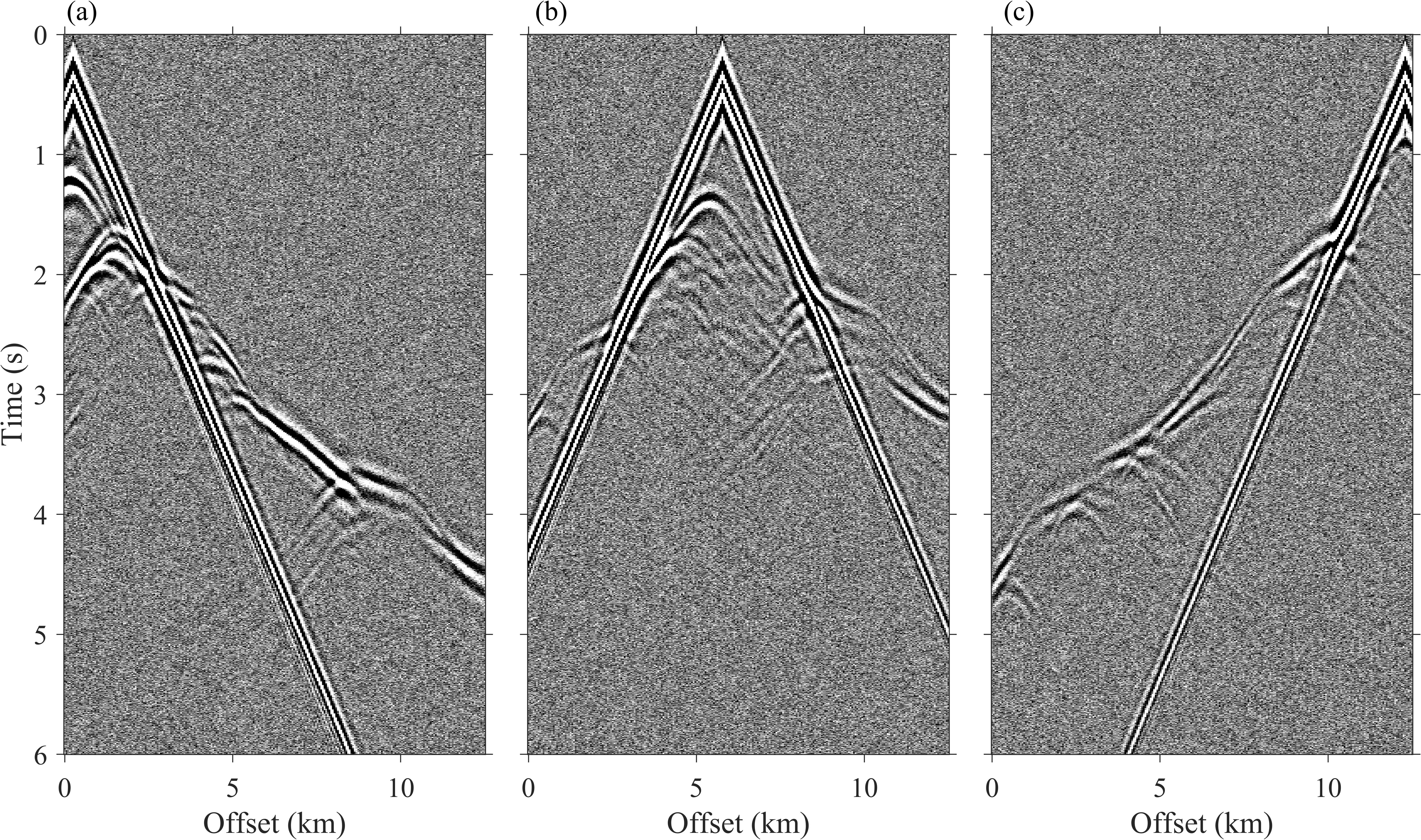}
\caption{2004 BP noisy synthetic data. (a) The 1st shot. (b) The 12th shot. (c) The 25th shot.}
\label{./BP/dn_obs.jpg}
\end{figure}

\begin{figure}
\centering
\noindent\includegraphics[width=\textwidth]{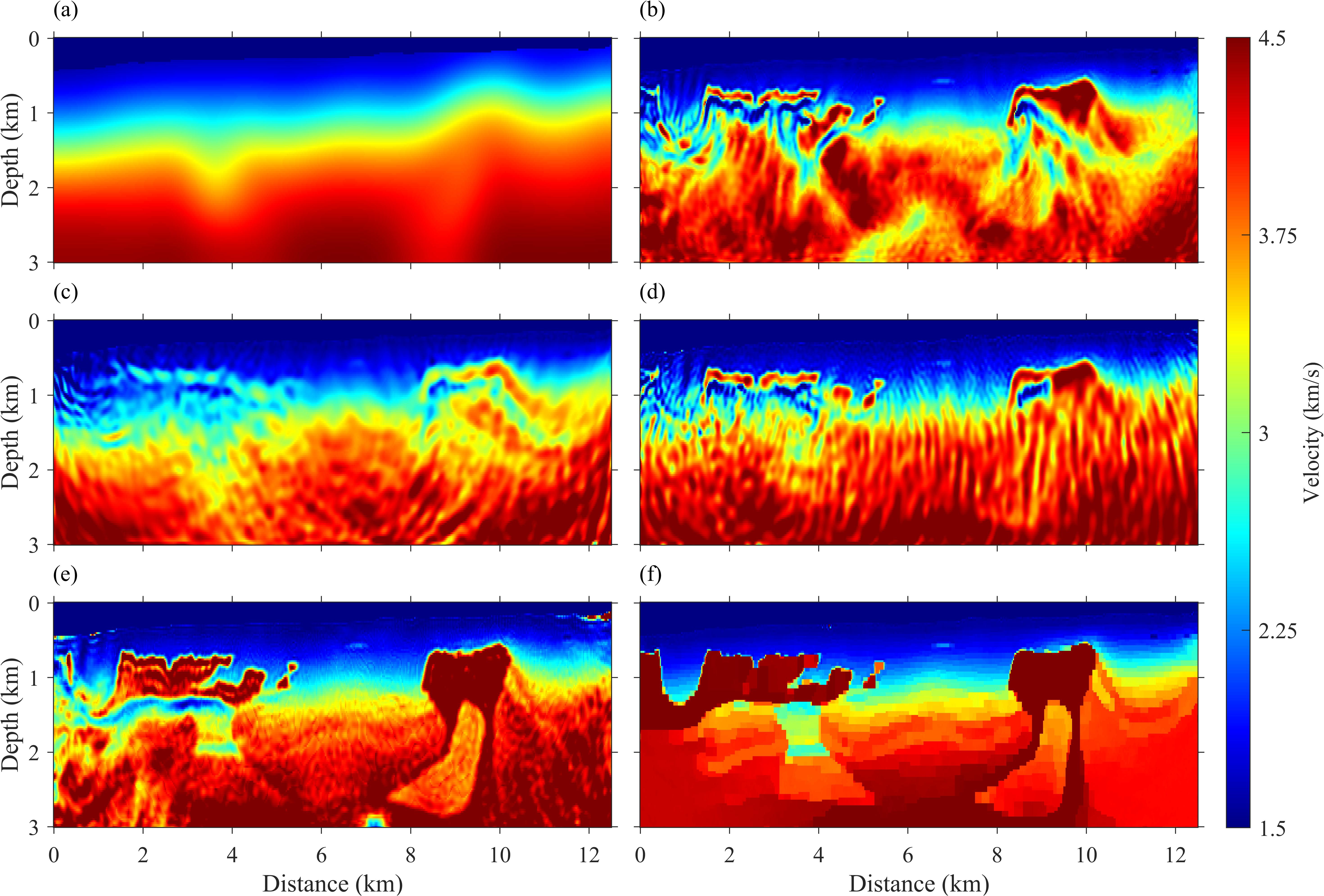}
\caption{2004 BP model with noise inversion results. (a) Initial model. (b) FWI. (c) EI. (d) W1-FWI. (e) MS-FWI. (f) MS{\_}ATpV-FWI.}
\label{./BP/BPn.jpg}
\end{figure}

To test the anti-noise ability of the proposed MS{\_}ATpV-FWI method, random noise with a signal-to-noise ratio of 10db is added to the synthetic seismic data corresponding to the 2004 BP model, as shown in Figure \ref{./BP/dn_obs.jpg}a-\ref{./BP/dn_obs.jpg}c. As observed, the waveform is contaminated by noise, increasing the difficulty of accurate waveform inversion.

Figures \ref{./BP/BPn.jpg}b-\ref{./BP/BPn.jpg}f present the inversion results for FWI, EI, W1-FWI, MS-FWI, and MS{\_}ATpV-FWI using noisy seismic data. In the presence of noise, FWI, EI, and W1-FWI become unstable, with their ability to accurately depict the salt dome structures significantly weakened. In contrast, the proposed MS{\_}ATpV-FWI method still produces high-resolution and accurate inversion results, demonstrating its superior anti-noise ability.

Table \ref{RS4} shows the RMSE and SSIM of the inversion results in the noisy case. The performance of other methods has degraded significantly, while the performance of the proposed method MS{\_}ATpV-FWI remains stable.

\begin{table}
\caption{RMSE and SSIM of the 2004 BP model with noise inversion results.}
\label{RS4}
    \centering
    \begin{tabularx}{\textwidth}{c>{\centering\arraybackslash}X>{\centering\arraybackslash}X>{\centering\arraybackslash}X>{\centering\arraybackslash}Xc}
    \hline
    \multirow{1}{*}{{Method}} & \multicolumn{1}{c}{FWI} & \multicolumn{1}{c}{EI} & \multicolumn{1}{c}{W1-FWI} & \multicolumn{1}{c}{MS-FWI} & \multicolumn{1}{c}{MS{\_}ATpV-FWI}\\
    \hline
    RMSE & 634.61 & 638.18 & 487.93 & 389.69 & \bf{185.39} \\
    SSIM & 0.44 & 0.42 & 0.45 & 0.68 & \bf{0.85} \\
    \hline
    \end{tabularx}
\end{table}

\subsection{Application on Field Seismic Data}

\begin{figure}
\centering
\noindent\includegraphics[width=\textwidth]{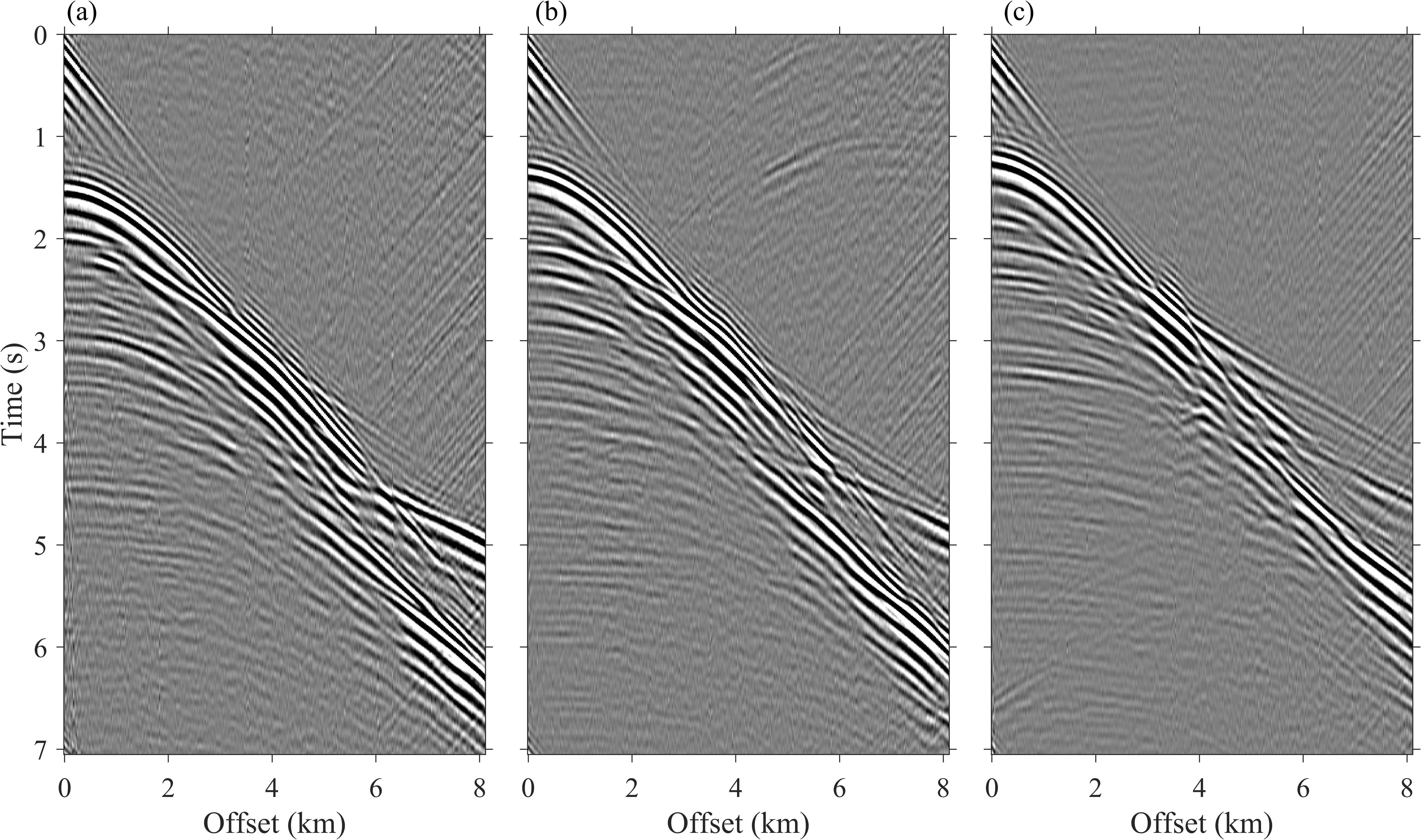}
\caption{Field seismic data. (a) The 1st shot. (b) The 50th shot. (c) The 100th shot.}
\label{./Field/d_obs.jpg}
\end{figure}

\begin{figure}
\centering
\noindent\includegraphics[width=3.1in]{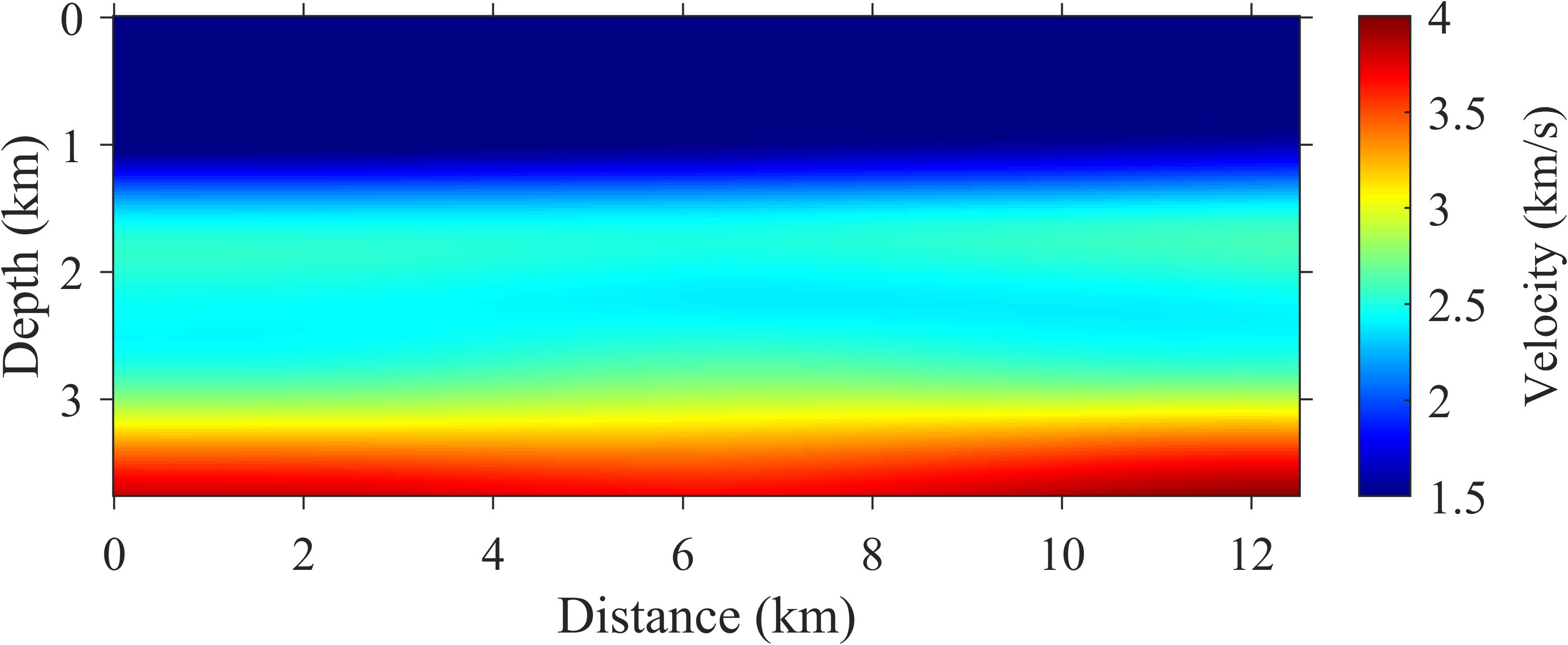}
\caption{The initial model for field data inversion.}
\label{./Field/Int.jpg}
\end{figure}

Finally, we test the performance of the proposed method MS{\_}ATpV-FWI on field data acquired over the North-Western Australia Continental Shelf by Viridien (formerly CGG). We selecte 100 shot gathers from the original data to cover our study survey area. The study area spatial grid size is $150\times500$, extending 12.5 km laterally and 3.75 km in depth, with a spatial sampling interval of 25 m. Figures \ref{./Field/d_obs.jpg} and \ref{./Field/Int.jpg} show the field seismic data acquired from the streamer and initial model for inversion, respectively.

\begin{figure}
\centering
\noindent\includegraphics[width=\textwidth]{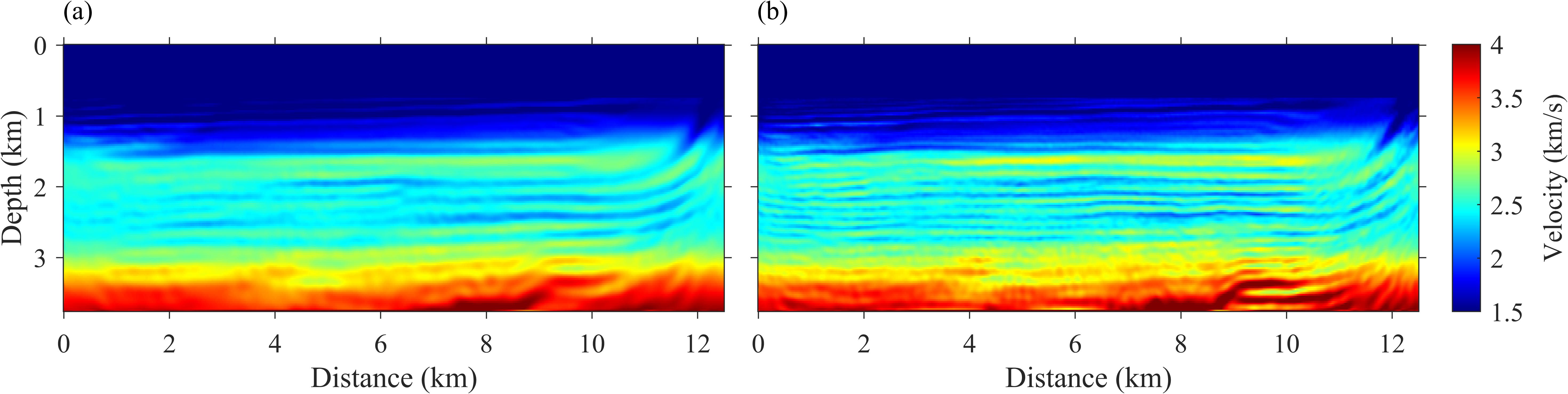}
\caption{Field seismic data inversion results. (a) 2-8 Hz. (b) 2-12 Hz.}
\label{./Field/result.jpg}
\end{figure}

\begin{figure}[htbp]
    \centering
    \begin{subfigure}[b]{\textwidth}
        \centering
        \includegraphics[width=1.0\textwidth]{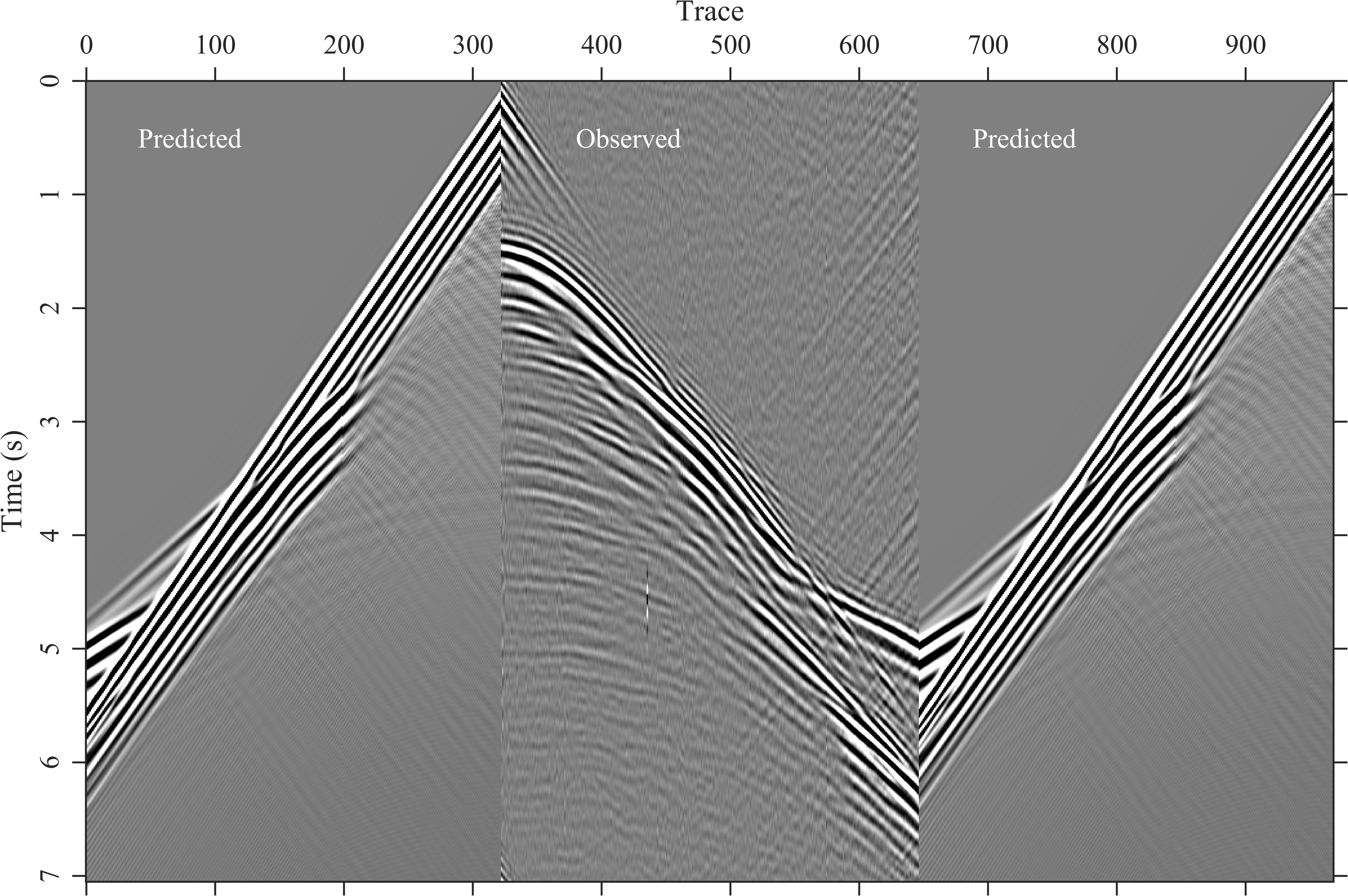}
        \caption{}
        \label{fig:sub1}
    \end{subfigure}
    \hfill
    \begin{subfigure}[b]{\textwidth}
        \centering
        \includegraphics[width=1.0\textwidth]{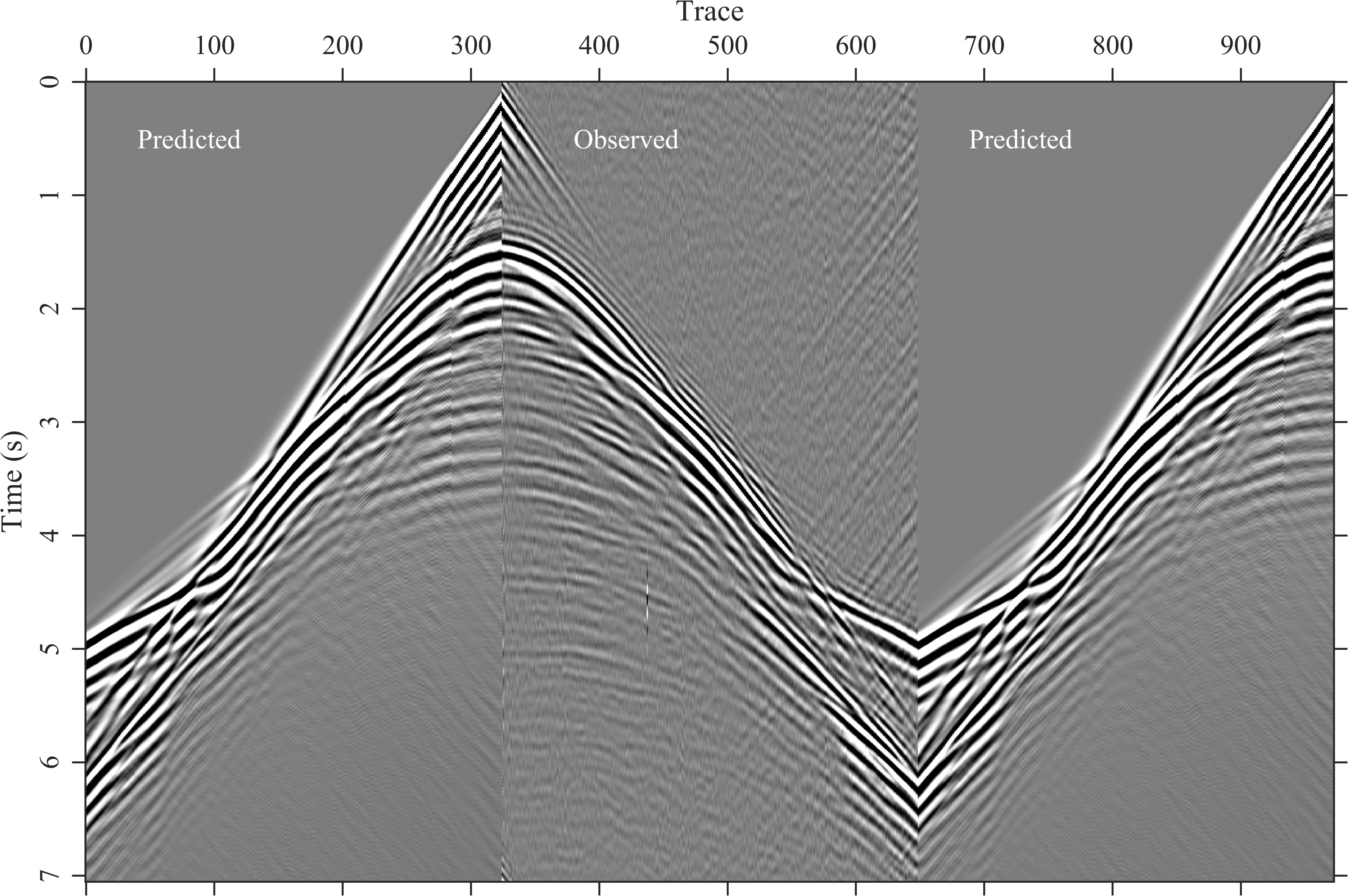}
        \caption{}
        \label{fig:sub2}
    \end{subfigure}
    \caption{Comparison of the predicted data generated by (a) the initial model, (b) the inverted model obtained by MS{\_}ATpV-FWI with the observed data.}
    \label{./Field/d_oi.jpg,./Field/d_op.jpg}
\end{figure}

It can be observed that the low-frequency components (below 2.5 Hz) in the field seismic data are contaminated by noise, which can negatively impact the accuracy and resolution of the inversion results. For this field data test, we employ a two-stage inversion strategy. In the first stage, we use bandpass-filtered seismic data in the frequency range of 2-8 Hz to update the velocity model. The MS{\_}ATpV-FWI inversion result is shown in Figure \ref{./Field/result.jpg}a. The MS{\_}ATpV-FWI method provides inversion result with rich velocity structure details and good horizontal continuity, demonstrating its effectiveness in the field data application. In the second stage, we further refine the details of the velocity model using bandpass-filtered seismic data in the frequency range of 2-12 Hz. The result is shown in Figure \ref{./Field/result.jpg}b. With the inclusion of higher-frequency seismic data, the inversion result is able to capture more detailed thin-layer information.

\begin{figure}
\centering
\noindent\includegraphics[width=5in]{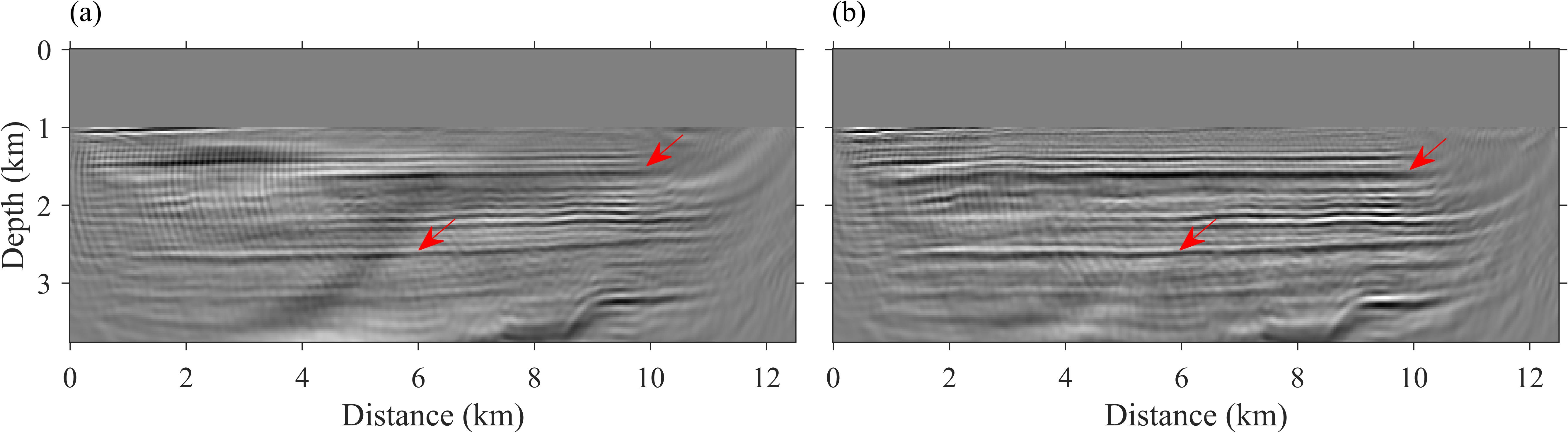}
\caption{LSRTM images from (a) the initial model and (b) the inverted model obtained by MS{\_}ATpV-FWI.}
\label{./Field/rtm.jpg}
\end{figure}

To thoroughly assess the reliability of the inversion result, we compare the predicted data generated from the initial model and the inversion model with the observed data, as shown in Figures \ref{./Field/d_oi.jpg,./Field/d_op.jpg}a and \ref{./Field/d_oi.jpg,./Field/d_op.jpg}b, respectively. The comparison demonstrates that the predicted data corresponding to the initial velocity model does not fit the observed data in the diving waves at the far offset and lacks reflections. While the predicted data from the inverted model more closely fits the observed seismic data  in both diving waves at the far offset and reflections at the near offset, indicating the accuracy of the inverted velocity from the proposed method. Next, we generate least-squares reverse time migration (LSRTM) results using both the initial and inverted models, as shown in Figures \ref{./Field/rtm.jpg}a and \ref{./Field/rtm.jpg}b, respectively. Due to velocity errors, the LSRTM image from the initial model exhibits distorted and discontinuous seismic events. In contrast, the LSRTM image obtained from the inverted model displays better continuity and energy focusing, which indicates that the inverted velocity model is kinematically correct.

\begin{figure}
\centering
\noindent\includegraphics[width=0.6\textwidth]{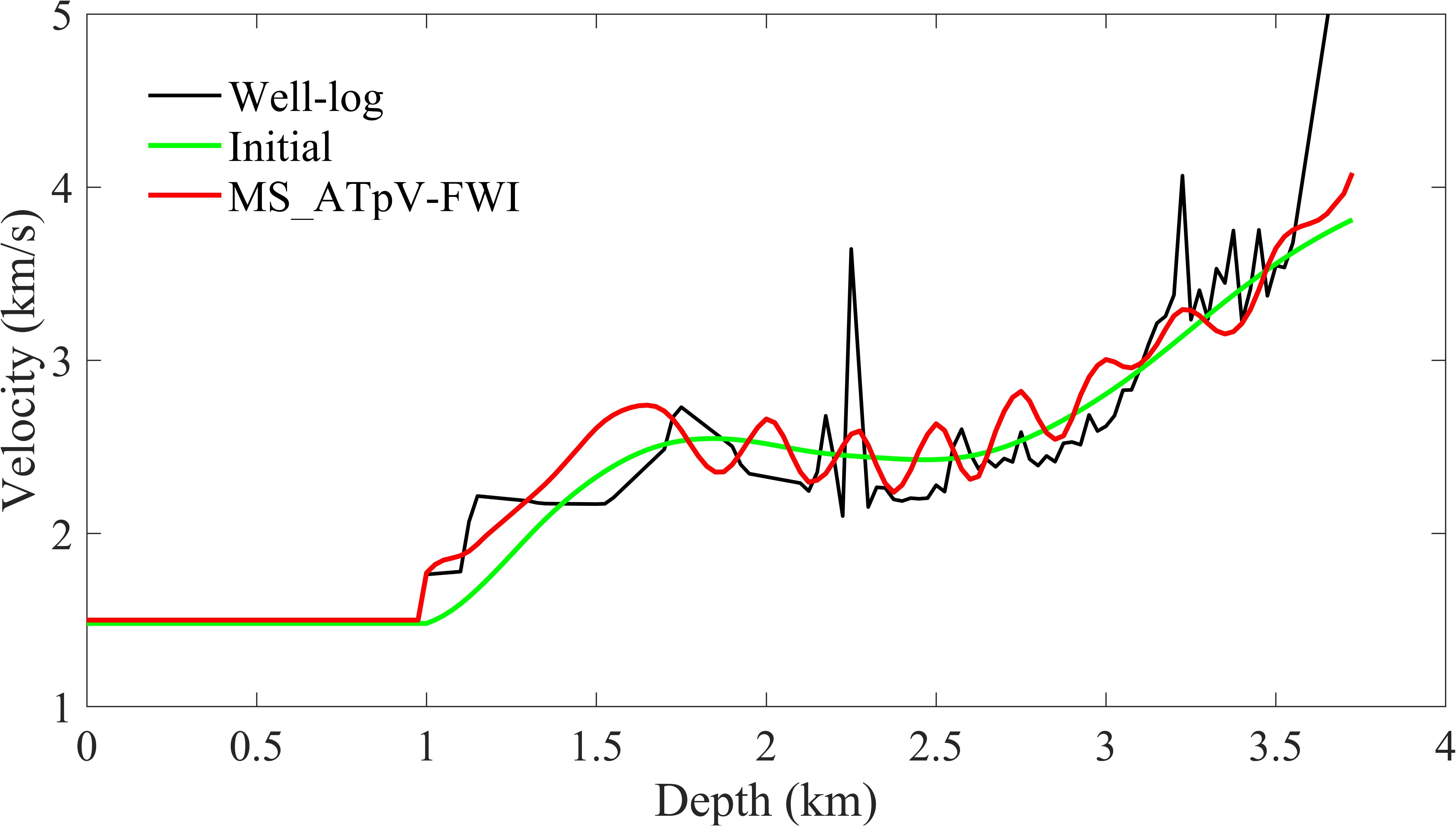}
\caption{Comparison of the vertical velocity profiles at location of 10.5 km.}
\label{./Field/well-log.jpg}
\end{figure}

To further validate the accuracy of the proposed MS{\_}ATpV-FWI method, we extract the curve from the inverted velocity at a horizontal position of 10.5 km and compare it with the well-log velocity, as shown in Figure \ref{./Field/well-log.jpg}. The inversion result from MS{\_}ATpV-FWI aligns most reasonably well with the well-log velocity, demonstrating the accuracy and practical applicability of the proposed method.

\section{Conclusion}

In this paper, we propose a full waveform inversion (FWI) method that combines multi-scale structural similarity index measure (M-SSIM) and anisotropic total variation regularization with the $l_p$ quasi-norm (ATpV). This method enables high-precision and high-resolution velocity inversion. By employing the M-SSIM, the method constructs an objective function that extracts multi-scale structural features in both time and space directions. This multiscale structural information helps to reduce the risk of cycle skipping and improves the accuracy of FWI. Additionally, the proposed method uses ATpV regularization to impose structural constraints on velocity gradients, suppressing artifacts in the gradients and enhancing the vertical resolution and lateral continuity of the inverted velocity model. The objective function is optimized efficiently and stably using the automatic differentiation (AD) technology, which reduces computational complexity. The effectiveness of the proposed method is demonstrated through both synthetic and field seismic data. Results show that the proposed method outperforms other popular FWI techniques, particularly when the initial model is inaccurate, the seismic data lacks low-frequency information, or contains noise, achieving accurate subsurface characterization of complex velocity structures.




\acknowledgments

We thank the editor, associate editor, and reviewers for their constructive suggestions, which greatly improved the manuscript. This work was supported by the National Key Research and Development Program of China under Grant (2023YFC3707901).

%
%

\bibliography{FWI}

\end{document}